\input harvmac
\def \E{{\cal E}}
\def \tt {\tau}

\def \arctan {{\rm arctan}}

\def \four{{\textstyle {1\ov 4}}}
  
\def \bI { {\bf 1}}
\def \rF {{\rm F}}
\def \ep{\epsilon}

\def \cN {{\cal N}}

\def \om {\omega}

\def \k {\kappa} 
\def \cF {{\cal F}}
\def \g {\gamma}
\def \del {\partial}

\def \const {{\rm const}}
\def \ha{{\textstyle{1\over 2}}}

\def \a {\alpha}
\def \b {\beta}
\def \chi {\chi}
\def \s {\sigma}

\def \m {\mu}
\def \n {\nu}
\def \vp {\varphi }

\def \t {\vartheta}
\def \t {\Theta}

\def \td {\tilde }
\def \d {\delta}

\def \inv {^{-1}}
\def \ov {\over }
\def \four{{\textstyle{1\over 4}}}
\def \fourth{{{1\over 4}}}

\def \hf {{\hat \vp}}
\def \t {\vartheta}

\def \f {{f^{(1)}_I}}
\def \ff {{f^{(2)}_I}}
\def \F {F^{(1)}}
\def \FF {F^{(2)}}
  \def \hf {\hat f} 
  \def \hF {\hat F}
\def \tb {{\tilde \vp} }
\def \diag  {{\rm diag}}
\def \s {{\rm  t}}
\def \rf{{\rm f}}

 \def \d {\delta} 
\def \L {{\Lambda}}

\def \hb {{\hat \vp}}

\def \lr { \lref}
\def\np {{  Nucl. Phys. }}
\def \pl {{  Phys. Lett. }}
\def \mpl {{ Mod. Phys. Lett. }}
\def \prl {{  Phys. Rev. Lett. }}
\def \pr  {{ Phys. Rev. }}

\def \cmp {{ Commun. Math. Phys. }}
\def \ijmp {{ Int. J. Mod. Phys. }}

\baselineskip8pt
\Title{
\vbox
{\baselineskip 6pt{\hbox{}}{\hbox
{Imperial/TP/97-98/27}}{\hbox{hep-th/9802133}} {\hbox{
  }}} }
{\vbox{\centerline { Open superstring partition function
 }
\vskip4pt
 \centerline { in  constant gauge field background  }
\vskip4pt
\centerline{ at  finite temperature   }
\vskip4pt
 \centerline { }}}\vskip -20 true pt
 \centerline{ A.A. Tseytlin\footnote{$^{\star}$}{\baselineskip8pt
e-mail address: tseytlin@ic.ac.uk}\footnote{$^{\dagger}$}{\baselineskip8pt
Also at Lebedev  Physics
Institute, Moscow.} }

\smallskip\smallskip
\centerline {\it Blackett Laboratory, 
Imperial College,  London,  SW7 2BZ, U.K. }

\bigskip\bigskip
\centerline {\bf Abstract}
\medskip
\baselineskip12pt
\noindent
We  find  the  general expression for  the  open superstring 
partition function  on the annulus in a constant abelian
gauge field background and at finite temperature. We  use the
approach based on Green-Schwarz string path integral in the 
light-cone gauge and compare it with NSR approach.
We  discuss the super Yang-Mills theory limit of the  string 
free energy  and mention 
some D-brane applications.

\Date {February 1998}

\noblackbox
\baselineskip 14pt plus 2pt minus 2pt
\lr\ft {E.S. Fradkin and A.A. Tseytlin, ``Non-linear electrodynamics from quantized strings", \pl B163 (1985) 123.   }
\lr\tse {A.A. Tseytlin, ``Vector field effective action in the open superstring theory", 
\np B276 (1986) 391; (E) B291 (1987) 876.}
\lr\tset {A.A. Tseytlin,
``Renormalization of Mobius infinities and                   
    partition function representation
 for string theory                       
    effective action",           \pl B202 (1988) 81. }

\lr\bat{I.A. Batalin, S.G. Matinyan and G.K. Savvidi, 
Yad. Fiz. 26 (1977) 407.}

\lr\baki{
C. Bachas and E. Kiritsis, {``$F^4$ terms in $N=4$ string vacua"}, 
\np Proc. Suppl. B55 (1997) 194, 
hep-th/9611205.}
\lr\kosh{A.L. Koshkarov and V.V. Nesterenko, 
``Open bosonic strings in a background isotropic electromagnetic field",
Dubna preprint, E2-89-555 (1989).}

\lr\dougl{M. Berkooz, M.R. Douglas and R.G. Leigh, 
``Branes Intersecting at Angles", \np B480 (1996) 265,
hep-th/9606139.}

\lr\zarem{
 I. Chepelev, Y. Makeenko  and K. Zarembo,
``Properties of D-Branes in Matrix Model of IIB Superstring", 
    Phys. Lett. B400 (1997) 43, hep-th/9701151.}

\lr\grr{M.B. Green, ``Point-like states for type IIB superstrings", 
\pl B329 (1994) 435.}
\lr\mac{M. McGuigan, ``Finite-temperature string theory and twisted tori", 
\pr D38 (1988) 552.}

\lr\odin{A.A. Bytsenko, E. Elizalde, S.D. Odintsov
and S. Zerbini, 
``Laurent series representation for the open superstring free
energy", \pl B297 (1992)  275, 
hep-th/9209021.}

\lr\odn{I. Lichtsier,  A.A. Bytsenko and S.D. Odintsov,
``Thermodynamic properties of  open non-critical string in  
external electromagnetic field",
Acta Phys. Pol. B22 (1991) 761.}

\lr \odd{A.A. Bytsenko,  S.D.  Odintsov and L. Granda,
``One-loop free energy for D-branes in constant electromagnetic field",
Mod. Phys. Lett. A11 (1996) 2525.}

\lr \andr {O.D. Andreev and A.A. Tseytlin, 
``Partition function representation for the open superstring effective action: cancellation of M\"obius infinities and derivative corrections to Born-Infeld lagrangian", 
\np B311 (1988/89) 205;
 \pl B207 (1988) 157. }
\lr \mrt {R.R. Metsaev, M.A. Rahmanov and A.A. Tseytlin, 
``The Born-Infeld action as the effective action in the open superstring theory",
\pl B193 (1987) 207. }
\lr \grw{ D. Gross  and E. Witten, ``Superstring modifications of Einstein equations",  \np B277 (1986) 1. }
\lr \callan {A.A. Abouelsaood, C.G. Callan, C.R. Nappi and S.A. Yost,
``Open strings in background gauge fields", 
\np B280 (1987) 599. }
\lr \berg { E. Bergshoeff, E. Sezgin, C.N. Pope and P.K. Townsend,
\pl B188 (1987) 70.}
\lr \bergg { E. Bergshoeff, M. Rakowski and E. Sezgin, 
\pl B185 (1987) 371.}
\lr\frat {E.S. Fradkin and A.A. Tseytlin,
``Effective action approach to superstring theory", 
 \pl B160 (1985) 69. 
  }
\lr \leii{
R.G. Leigh, \mpl A4 (1989) 2767. }
\lr \lei{J. Dai, R.G. Leigh and J. Polchinski, 
\mpl A4 (1989) 2073.}
\lr \pol { J. Polchinski, ``Dirichlet branes and Ramond-Ramond charges",  Phys. Rev. Lett. 75 (1995) 
4724, hep-th/9510169;
``TASI lectures on D-branes", hep-th/9611050.
 }
\lr  \witt { E. Witten, \np B443 (1995) 85, hep-th/9510135. } 
\lr\appel{T. Appelquist and R.D. Pisarski, ``High temperature Yang-Mills theories and three-dimensional  quantum chromodynamics",
\pr D23 (1981) 2305.}

\lr \tst { A.A. Tseytlin,\np B469 (1996) 51,  hep-th/9602064. }
\lr \tseyy { Tseytlin  Int.J.Mod.Phys.A4:1257,1989.} 
\lr \scherk {A. Neveu and J.  Scherk, \np B36 (1972) 155;
J.  Scherk  and J.H. Schwarz, \np B81 (1974) 118. }

\lr \napp {P.C. Argyres and C.R. Nappi, \np B330 (1990) 151. }
\lr \kita {Y.  Kitazawa, \np  B289 (1987) 599.} 
\lr \prev { T. Hagiwara, J. Phys. A14 (1981) 3059.}
\lr \dorn {H. Dorn, hep-th/9612120; 
H. Dorn and H.-J. Otto, hep-th/9603186.}
\lr\ishi{
N. Ishibashi, H. Kawai, Y. Kitazawa and A. Tsuchiya, hep-th/9612115.}

\lr \anmpl {O.D. Andreev and A.A. Tseytlin, \mpl   A3 (1988) 1349.}

  \lr \fri{ C. Lovelace, \np B273 (1986) 413; 
  B.E. Fridling and A. Jevicki, \pl B174 (1986) 75.}
  
 \lr \hamada { K. Hamada, hep-th/9612234.}
  \lr \shif { M.A. Shifman, \np B173 (1980) 13.}
   \lr\nev {G.L. Gervais and A. Neveu, \np B163 (1980) 189; H. Dorn,
 Fortschr. d.  Phys. 34 (1986) 11.} 
 \lr\schw{M. Aganagic, C. Popescu and J.H. Schwarz, hep-th/9612080.}
 
\lr\nester{V.V. Nesterenko, ``The dynamics of open strings in a background electromagnetic field", 
 \ijmp A4 (1989) 2627.}

 \lr\brow{H. Leutwyler, \np B179 (1981) 129;
 L.S. Brown and W.I. Weisberger, \np B157 (1979) 285.}
 \lr \doug {M. Douglas, hep-th/9512077.}
 \lr \gre{M.B. Green and N. Seiberg, \np B299 (1988) 559.}
 
 \lr\lii{M. Li, hep-th/9612222.}
 
 \lr \grg{ M.B. Green and M. Gutperle, hep-th/9612127.} 

 \lr \grgu{ M.B. Green and M. Gutperle, ``Light-cone supersymmetry and D-branes", \np B476 (1996) 484,  hep-th/9604091.} 

\lr \bbpt{ K.~Becker, M.~Becker, J.~Polchinski, and A.A.~Tseytlin, 
{``Higher order graviton scattering in M(atrix) theory"}, 
Phys. Rev. D56 (1997) 3174, 
hep-th/9706072.}

\lr\ffi{E.J. Ferrer, E.S. Fradkin and V. de la Incera, 
``Effect of a background electric field on the Hagedorn temperature",
\pl B248 (1990) 281.}
\lr\pool{ J. Polchinski, ``Evaluation of the one loop string path integral", 
\cmp 104 (1986) 37.}
\lr\alo{E. Alvarez and M.A.R. Osorio, ``Superstrings at finite temperature",
\pr D36 (1987) 1175.}
\lr\vaz{M.A. Vazquez-Mozo, 
``Open string thermodynamics and D-branes", 
\pl B388 (1996) 494, hep-th/9607052.}
\lr\greee{M.B. Green, 
``Wilson-Polyakov loops for critical strings 
and superstrings at finite temperature",\np B381 (1992) 201.}
\lr\horpol{G.T. Horowitz and J.  Polchinski,
``A Correspondence Principle for Black Holes and Strings", 
hep-th/9612146.}

\lr\tayl{D. Kabat and W. Taylor, ``Spherical membranes in Matrix theory", 
hep-th/9711078;
``Linearized supergravity from Matrix theory", hep-th/9712185.}

\lr\atic{
J. Atick and E. Witten, ``The Hagedorn transition and the number of 
degrees of freedom of string theory", \np B310 (1988) 291.}

\lr\ahber{O. Aharony and M. Berkooz, hep-th/9611215.}
\lr \liff{G. Lifschytz, hep-th/9610125.}
\lr\tsey{A.A. Tseytlin, \prl 78 (1997) 1864,  hep-th/9612164.}
\lr\met{R.R. Metsaev  and A.A. Tseytlin, 
``One-loop corrections to string theory effective actions", 
\np B298 (1988) 109.}

\lr\berli{
O. Bergman, M. Gaberdiel and G. Lifschytz, 
``Branes, orientifolds and the creation of elementary strings", 
 \np B509 (1998) 194, hep-th/9705130.}

\lr\gree{ M.B. Green  and J.H. Schwarz, 
``Supersymmetric dual string theory. 3. Loops and renormalization", 
\np B198 (1982) 441.}

\lr\chep{I.~Chepelev and  A.A.~Tseytlin,
{``Long-distance interactions of branes: correspondence  between 
supergravity  and super Yang-Mills descriptions"},
hep-th/9709087.}

\lr\dkps{  M.R. Douglas, D. Kabat, P. Pouliot and S.H. Shenker,
``D-branes and short distances in string theory",  \np B485 (1997) 85,
hep-th/9608024. } 

\lr\chets{I. Chepelev and A.A. Tseytlin, ``Interactions of type IIB D-branes from D-instanton matrix model", 
hep-th/9705120.}
\lr \bacp { C. Bachas and M. Porrati,
``Pair creation of open strings in an electric field", 
 \pl B296 (1992) 77.}
\lr\bacf{
C. Bachas and C. Fabre, ``Threshold effects in open string theory", 
hep-th/9605028.}

\lr \bac { C. Bachas, ``D-brane dynamics", \pl B374 (1996) 37, hep-th/9511043.}
\lr \lif{G. Lifschytz, ``Probing bound states of branes",
 Nucl. Phys. B499 (1997) 283, hep-th/9610125;
``Four-brane and Six-brane interactions in M(atrix) model", hep-th/9612223;
G. Lifschytz and S.D. Mathur, ``Supersymmetry and membrane interactions in M(atrix) theory",\np B499 (1997) 283,  hep-th/9612087.}

\lr\kiri{D. Mamford, ``Tata Lectures on Theta", Birkauser,  Boston, 1983;
E. Kiritsis, ``Introduction to superstring theory", 
hep-th/9709062, p. 212.}

\lr \frao { E.S. Fradkin and A.A. Tseytlin, 
``Quantum properties of higher dimensional and dimensionally reduced supersymmetric theories", 
\np B227 (1983) 252.
}
\lr \gsw { M. Green, J. Schwarz and E. Witten, 
``Superstring Theory", Cambridge U.P., 1987. }
\lr\call {C.G. Callan, C. Lovelace, C.R. Nappi and S.A. Yost, 
``Loop corrections to superstring equations of motion", 
\np B308 (1988) 221.}
\lr\cakl{C.G. Callan and I.R. Klebanov, ``D-brane boundary state dynamics", 
\np B465 (1996) 473, hep-th/9511173.}
\lr\li{M. Li, ``Boundary states of D-branes and Dy-strings", hep-th/9510161.}

\lr \elm{P. Elmfors and B.-S. Skagerstam, ``Electromagnetic
fields in a thermal background", \pl B348 (1995) 141, hep-th/9404106.}
\lr\seib{ 
M.B. Green  and N. Seiberg, 
 ``Contact interactions in superstring theory",
\np B299 (1988) 559.}
\lr \chap{S. Chapman, ``A new dimensionally reduced effective action for QCD at finite temperature", 
\pr D50 (1994) 5308, hep-ph/9407313.}

\lr\guk{S. Gukov, I.R. Klebanov and A.M. Polyakov, 
``Dynamics of (n,1) strings", hep-th/9711112.}
\lr \TT{A.A. Tseytlin, ``Sigma model
 approach to string theory",                    
    Int. J. Mod. Phys. A4 (1989) 1257;
 Int. J. Mod. Phys. A4 (1989) 4249.}

\lr \green {M.B. Green and J.H. Schwarz, 
``Anomaly cancellations in supersymmetric D=10 gauge theory require SO(32)", 
\pl B149 (1984) 117; ``Infinity cancellations in SO(32) superstring theory", 
\pl B151 (1985) 21.    }
\lr \bgs {M.B. Green,  J.H. Schwarz  and L.  Brink, 
``N=4 Yang-Mills and N=8 supergravity as limits of string theories", 
\np B198 (1982) 474.  }

\lr\bill{M. Billo, P. Di Vecchia, M. Frau, A. Lerda, I. Pesando, R. Russo and S. Sciuto, 
``Microscopic string analysis of the D0-D8 brane system and dual R-R states", 
hep-th/9802088.}

\lr\pierre{ J.H. Pierre, ``Interactions of eight-branes in string theory and matrix theory", Phys. Rev. D57 (1997) 1250,  hep-th/9705110.      } 

\lr \lerch {W. Lerche, ``Elliptic index and superstring effective actions", \np B308 (1988) 102.}

\lr\tsr{J.G.  Russo and A.A. Tseytlin, ``Magnetic flux tube models in superstring theory", \np B461 (1996) 131, hep-th/9508068.}

\lr\sheih{M.M. Sheikh Jabbari, ``Classification of different branes at angles",  hep-th/9710121.}
\lr\mald{J. Maldacena, ``Probing near extremal black holes with D-branes", 
hep-th/9705053.}
\lr\calm{C.G. Callan and J. Madacena, ``D-brane Approach to Black Hole Quantum Mechanics", Nucl. Phys. B472 (1996) 591,  hep-th/9602043.      }

\lr\igor{ S.S. Gubser, I.R.  Klebanov and A.W. Peet, 
``Entropy and temperature of black 3-branes", 
\pr D54 (1996) 3915, hep-th/9602135.}

\lr \li{M. Li,  ``Boundary States of D-Branes and Dy-Strings", 
Nucl. Phys. B460 (1996) 351, hep-th/9510161;
M. Douglas, ``Branes within Branes", 
hep-th/9512077.
 }

\lr\cabo{A. Cabo, O.K. Kalashnikov and A.E. Shabad, ``Finite temperature gluonic gas in a magnetic field", \np B185 (1981) 473.}
\lr\nilol{N.K. Nielsen and P. Olesen,
``An unstable Yang-Mills field mode",
 \np B144 (1978) 376;
V.V. Skalozub, ``On restoration of spontaneously 
broken symmetry in magnetic field", Yad. Fiz. 28 (1978) 228.}

\lr\burg{C.P. Burgess, ``Open string instability in background electric fields", 
\np B294 (1987) 427.}
\lr\bank{T. Banks, W. Fischler, S. Shenker and L. Susskind, 
``M-theory as a matrix model: a conjecture", 
\pr D55 (1997) 5112, hep-th/9610043.}
\lr\mas{J. Maldacena and A. Strominger, ``Black hole greybody factors and D-brane spectroscopy",  Phys. Rev. D55 (1997) 861,  hep-th/9609026.}

\newsec{INTRODUCTION}
Leading-order interactions between BPS states of 
 D-branes \pol\ admit both supergravity
and super Yang-Mills descriptions  which give equivalent 
results for interaction potentials (see, e.g., \refs{\bac,\dkps,\bank,\lif,\chep}  
and references there). This equivalence can be understood as being, in particular, 
  a consequence of  the 
universality of the  leading $F^4$ term in the  open superstring partition 
function on the annulus \refs{\dkps,\baki}.

An important question is whether this correspondence 
between the two descriptions  applies also 
to non-BPS (excited, or non-extremal)  states of D-branes.
As was found  in \mald,  
starting with the one-loop  $F^4$-term in the SYM effective action
and assuming  certain averaging over  SYM backgrounds which have the right  
energy and charges to describe  near-extremal  branes on supergravity side
one  obtains the expressions   which  
have the same structure  as  supergravity
 interaction potentials between extremal and near-extremal branes. 
The precise coefficients do not seem to  match  
however.\foot{The agreement for potentials between non-BPS branes 
found in \tayl\ seems to apply 
only to  a  class of  configurations which are spherically symmetric
 in $D=11$ sense.}

From   supergravity  point of view, 
 the  non-extremal branes  can be assigned certain temperature
 and entropy.
It is thus  natural to expect that the SYM description of non-extremal 
  RR branes should, in fact,  be based on {\it thermal} gauge theory  states. 
 In particular, one may try to interpret   the Hawking radiation 
of a certain class of near-extremal  black holes with RR charges
in terms of emission of closed string modes  by a D-brane configuration 
in an excited state  \refs{\calm,\mas}.
The  entropy of near-extremal D3-brane can be reproduced 
as the entropy of finite-temperature  ensemble of  states of  the $N=4$ 
four-dimensional 
SYM  theory    \igor.
As discussed in \horpol,  there are 
two kinds of non-extremal    states 
of D-branes: one can be thought of as a D-brane
with a small number of long (massive)  strings, 
and another  as a D-brane  with  a large number of short  (massless) 
open strings. For large  deviation from extremality (or large temperature, \ 
 $\b <  \sqrt{\a'}$)
 long string state  has greater entropy, while for small 
excess  energy  the gas of light open strings
(or SYM modes)  is the relevant  description.

This suggests that to describe the potential between 
non-extremal D-branes  one should  perform   finite-temperature 
analogs  of computations in \refs{\bac,\dkps,\lif,\chets},
i.e. determine  the  corresponding  terms in the  finite-temperature 
open string partition function  or   finite-temperature 
effective action of SYM theory.

This is one  of  motivations behind the formal discussion
 of the present paper.
In  more general context, it is of interest to study the 
combined effect of the temperature and magnetic and electric 
background fields on the behaviour of  open string  ensemble.
Here we  shall compute the finite temperature  superstring partition function 
on the annulus in  a  constant gauge field background and  
obtain the corresponding SYM free energy in the $\a'\to 0$ limit. 
Possible applications will be mentioned only briefly. 

We shall start in Section 2 with a detailed discussion 
of the zero-temperature case. 
The  string partition function  in  external fields 
is directly related to the  string effective action.\foot{Low-energy  effective action in string theory 
can be defined as  a `superposition' of 
string scattering amplitudes with massless tree-level poles subtracted. 
The scattering amplitudes are given  by correlators of 
 vertex operators in flat background. 
In the Polyakov path integral approach 
it is possible to represent the string effective action 
in terms of the (renormalized) 
 string  sigma model  partition function  in background fields. Indeed,  
the latter is the generating functional for string amplitudes
(average of the exponential of vertex operators multiplied by external fields) and renormalisation of logarithmic 2d divergences 
 effectively  subtracts the massless poles 
 \refs{\frat,\TT}.}
This relation is  particularly simple  
in the open string theory case.\foot{At the tree (disc) level the divergences associated with 
  the M\"obius group volume  are absent in the superstring case \andr\
and  can be easily renormalised away in the bosonic case \refs{\ft, \tse}. 
 The problem of M\"obius infinities does not appear at one and 
higher loop level
where  in computing $Z$ 
 one needs only to subtract logarithmic divergences  associated with  
 massless poles  in the  amplitudes. 
The local part of these divergences is absent in the case
 of the constant abelian  vector field  strength
background   so that $Z(F)$ is finite
(apart from modular divergences that may or may not cancel depending on a particular problem and theory under consideration).} 
The computation of the  one-loop (annulus) 
superstring partition function in a constant gauge field background 
$Z(F)$ is  straightforward
 in the light-cone gauge Green-Schwarz (GS)  formulation 
 and was  originally  considered  in \tse. 
In section 2.1  we shall put $Z(F)$ found  in \tse\ in  a more   explicit
form and then in section 2.2  compare it with  related results 
found using NSR path integral 
and GS boundary state approaches.\foot{One of our  aims is to  clarify the structure of the light-cone GS  
path integral approach
to computation of the string partition function  
with a hope that it can be  applied to 
the problems of determining   derivative
$O(\del F)$  corrections to the one-loop  $Z(F)$ 
and  computing the two-loop contribution  to the 
string effective action
(in particular, in order to check 
 the conjecture \refs{\chep,\bbpt}
that, like $F^4$ term in one-loop $Z$ \dkps, 
 the $F^6$ term in two-loop $Z$ is `universal', i.e. 
has trivial dependence on $\a'$ and thus on space-time
 IR cutoff or distance between branes). 
Such  computations 
in the NSR formalism  where one needs to sum 
over spin structures  appear  to be very  complicated.
Related 
examples of the utility  of the l.c. GS approach to computing  the one-loop
partition function in closed string theory can be found in \refs{\lerch,\tsr}. 
} 

For constant  $F$ the  GS path integral becomes gaussian 
in both bosonic ($x$) and fermionic $(S)$ coordinates.
The approach used in \tse\ was to compute the corresponding 
determinants with the free-theory boundary conditions
$\del_n x=0, \  S_1=S_2$. This is in direct correspondence with 
the standard  definition of the string scattering amplitudes 
as correlators of l.c. gauge GS vertex operators \refs{\gree,\green,\gsw}.
Somewhat surprisingly, we find that the resulting  $Z(F)$
 is  different   from the result of the NSR  approach of \bac\ 
(which is  equivalent to the result of the GS boundary state approach 
of \grgu).
The two partition functions agree in the  $\a'\to 0$  (massless state)
limit  but differ in  the  contributions of massive states.

Similar gaussian path integral in the NSR approach  can be computed either 
directly,  by evaluating determinants in external field $F$ 
using free-theory boundary conditions, or, as in \bacp, by 
first solving  the  boundary conformal field theory  
in terms of free oscillators (imposing  
 $F$-dependent boundary conditions
 \refs{\callan,\call})  and then  obtaining  the 
partition function using  the  standard  operator formalism relation
  (the 
two procedures give, of course,  equivalent expressions   for $Z(F)$).
The same result can be found   in the GS approach
if instead of computing  path integral with the free-theory boundary conditions 
as in \tse\ one first solves  the  2d  conformal theory
using   appropriate (`supersymmetric')  boundary condition 
$S^a_1 = \hat M^{ab}(F)  S^b_2$  \grgu. This boundary condition 
  is  the one that is consistent 
with space-time supersymmetry \refs{\call,\grgu}  (though
  it is  not the one  implied by 
the form of the $F$-dependent l.c. gauge GS action). 
 
 While  the approach of \tse\ 
is in direct correspondence with the standard l.c. gauge GS vertex operator 
definition of the  string  amplitudes, it may not  manifestly 
preserve  supersymmetry.\foot{Indeed, here one uses the 
 constant $F$ truncation,
i.e. keeps only  (momentum)$^n$ term 
in the $n$-point vector amplitude
 while  the symmetries of the theory are expected 
to  be preserved provided one 
defines the amplitudes by an analytic continuation in momenta.} 
This  suggests that the  difference between the expressions  of \tse\
and \refs{\bacp,\grgu} may  be interpreted 
as being due to  certain  `contact terms' 
 \seib\  (which for constant $F$  may   have 
the same  structure  as  the `main' terms). 
The two expressions  may be  related by a 
redefinition of the field strength 
$F\to F + O(\a' F^2)$ containing all powers of $F$.\foot{A lesson seems to be 
that one   may get inequivalent results for a superstring partition function 
depending  on  whether one reconstructs it from an expansion 
near free-theory  point or  defines  it in terms 
of  a non-trivial conformal  field theory  which uses 
appropriate boundary conditions.}

The comparison with the approach of \grgu\
shows how  the GS path integral result of \tse\
 is  to be modified  to restore its equivalence
with the NSR result. 
In section 2.2 we shall first  find  $Z(F)$ 
depending on  four `magnetic' eigenvalues of the $D=10$ background $F_{\m\n}$
and then generalise it to include the dependence on the fifth 
 `electric' component, obtaining
the  $D=10$ Lorentz-invariant expression for the partition function
(related partial results appeared in \refs{\lif,\pierre,\sheih}).
 Some properties of $Z(F)$ and D-brane applications
will be discussed in section 2.3.
In section 2.4 we shall  show that  the $\a'\to 0$ limit of $Z(F)$ 
is equivalent to  the one-loop  SYM effective action
in a general $D=10$ constant abelian gauge field background.

The finite temperature  case 
will be the subject of section 3. 
In section 3.1 we shall find the finite-temperature analogue
of the one-loop  effective action (free energy) in SYM theory in a magnetic 
background. 
In section 3.2 we shall use the GS approach to 
generalise the string partition function 
obtained in section 2.2 to the finite-temperature case.
The resulting expression $Z(\b,F)$ will have the SYM free
 energy as its $\a'\to 0$ limit.
In section 3.3 
 we shall study the  dependence  of $Z(\b,F)$ 
on the background field 
and   temperature.  In particular, we shall show that the value 
of the Hagedorn temperature is not modified by the magnetic field
and that the  infra-red  magnetic instability 
of zero-temperature  partition function 
 (present in both  string theory 
and  SYM theory)  remains  also at finite temperature.
The finite-temperature case with  an electric 
background  will be  discussed   in section 3.4. 
The presence of an  electric field  is known  to 
modify the  value of the Hagedorn temperature of the neutral open  bosonic string gas 
\ffi\ and  we find that the same is true for the 
open superstring gas.

\newsec{OPEN SUPERSTRING PARTITION FUNCTION ON THE ANNULUS 
IN  CONSTANT BACKGOUND FIELD (ZERO TEMPERATURE CASE)}
\subsec{\bf   Green-Schwarz  path integral with free-theory
  boundary conditions}

Our starting point is the open superstring partition function 
in the constant abelian gauge field background
given by light-cone gauge GS path integral 
 on the annulus\foot{We shall 
be interested in the case of oriented open superstrings  which is relevant for the description of D-branes and  discuss only the annulus diagram. All considerations
below can be straightforwardly repeated for the  case of the M\"obius strip diagram of type I theory.}  \refs{\frat,\tse}
\eqn\gen
{Z (F^{(1)}, F^{(2)}) = 
\int [d q] [dx][dS] \  \exp [i(I_0 + I^{(1)}_{int} + I^{(2)}_{int}    )]  \ , } 
where (the indices $i,j$ and $a,b$ run from 1 to 8)
$$
I_0 = - { 1 \ov 4 \pi \a' } 
\int d^2\sigma\  [ \del_+ x^i \del_- x^i -
 {i \ov 2 } \a' ( S^a_1 \del_+ S^a_1 + S^a_2 \del_- S^a_2)  ]  \ , 
$$
\eqn\ope{
I_{int} = \int dt\  [ \dot x^i A_i (x)  - {i \ov 2} \a' 
S^a  S^b\hat  F_{ab} ]
\ ,  }
\eqn\yyy{
\hat F_{ab} = \four \g^{ij}_{ab} F_{ij}\ .  }
For  the constant field 
strength,   \eqn\oppo{A_i= - \ha F_{ij} x^j\  , \ \ \ \ \ \ \  \ F_{ij}=\const \ , } 
 the  path integral 
is gaussian \refs{\ft,\tse}.
To be able to  use the l.c. gauge  GS formalism, 
we assume that the electric field
components are vanishing. 
At zero temperature  the final result for $Z(F)$  should admit  a Lorentz-covariant
 $SO(1, 9)$
generalisation,  so that 
 the dependence on  electric components  
may be  deduced by  analytic  continuation in magnetic components.

We shall  assume the    standard \gsw\ 
boundary conditions
 \eqn\bccs{
\del_\sigma x^i=0\ , \ \ \ \ \   S^a_1=S^a_2 \equiv S^a \ , }
with both $x$  and $S$ being 
 periodic in  the angular coordinate of the annulus.    For generic  vector fields 
$Z$ is the generating functional for  the scattering amplitudes
($I_{int}$ corresponds to the standard l.c. gauge form of the vector 
field vertex operator \gsw).
The form of  the interaction action \ope,\yyy\
follows also  upon fixing the l.c. gauge in the  covariant 
action for a  GS superstring coupled to the on-shell $D=10$ SYM 
 superfield background \frat.

 In the present case of  a  gaussian
theory  one may  compute the path integral either by  first 
directly expanding the fields subject to the free-theory boundary conditions
\bccs\ 
 in modes and then integrating them out, or by 
first solving the classical equations of the theory  using 
 field-dependent 
boundary conditions implied by   the  action  \ope, i.e.
\eqn\som{   
\del_\sigma x^i + F^{ij} \del_\tau x^j =0\ ,   \ \ \ {\rm or} \ \ \ \
 \del_+ x^i = M^{ij} \del_- x^j \ ,  \ \ \ \ \ 
M= { \bI +  F \ov \bI -  F} \ , }
\eqn\iop{S^a_1 - S^a_2 = \hat F^{ab} (S^b_1 + S^b_2 ) \ , \ \ \ {\rm or} \ \ \ \
S^a_1 = \td  M^{ab} S^b_2\ , \ \ \  \
 \ \td M = { \bI + \hat F \ov \bI - \hat F} \ . }
The results for $Z(F)$ 
found using  these two  approaches are, of course, equivalent.

One may, in principle, consider a possibility of replacing 
$\hat F_{ab}$ in \yyy\    with
 $\hat F_{ab}'$ given by a local power series in $F$, \ 
$\hat F_{ab} = \four \g^{ij}_{ab} F_{ij} +  O(F^2), $ as this 
would modify the corresponding scattering  amplitudes only by certain  
contact terms
(which may be necessary to add  to  maintain Lorentz symmetry or 
space-time supersymmetry \seib). Equivalently, one 
may   consider replacing \iop\ by a different  field-dependent 
boundary condition  (implied by the action \ope\ with $\hat F$ 
given not  by \yyy\ but by $\hat F_{ab}$) 
\eqn\nnn{ S^a_1=\hat M^{ab} S^b_2  \ , \ \ \ \ \ \ \ \ \ 
   \hat M = \hat M (F) \ .  }
This again  would change the  scattering amplitudes  only by contact terms.
The choice of such modified boundary condition  with 
$\hat M^{ab}$ related to $M$ in \som\  as a spinor rotation
is related  to a vector rotation, i.e.   by 
\eqn\yyyc{
M^{ij} \g^j = \hat M^{-1} \g^i \hat M  \ ,    }
follows from 
 the  condition  that the boundary conformal field  theory defined by 
\ope,\oppo\ 
should  preserve  space-time supersymmetry \refs{\call,\grgu}.
As we shall find below,  such  modification is, indeed, 
required 
in order to obtain the same expression  for $Z$ as 
in manifestly Lorentz-covariant
NSR path integral approach 
(in the  special  case  of $F$ having only one non-vanishing component
 this was  shown in 
\grgu\ where the NSR result of \bacp\ was reproduced using \yyyc). 

In  this subsection we shall follow  the  
approach of \tse\  based on \yyy,\bccs\ but   many  of  equations   
below  will not depend on  the  explicit form of the 
 relation between $\hat F$ and $F$ so  that the 
modification required for restoring the   equivalence 
 with the NSR approach 
  will be straightforward to implement later. 
We shall  often absorb $2\pi \a'$ in the two  magnetic
 field 8$\times$8  matrices $F^{(r)}_{ij}$  ($r=1,2$) 
representing the interactions at the  boundaries of the annulus.\foot{In the case of the open string with charges $e_1,e_2$ in an external  magnetic field
 $ F_{ij}$ one has $F^{(r)}_{ij} = e_r F_{ij}$
(the neutral string case  corresponds to 
$e_1 +e_2 =0$  or $F^{(1)}_{ij} + F^{(2)}_{ij} =0$).
We shall, however, keep  $F^{(1)}_{ij} $ and $  F^{(2)}_{ij}$
 independent  which is useful in view of D-brane applications.}
The  gaussian path integral over  the  bosons and fermions 
gives \tse
\eqn\zet{
Z= c_0  \ 
\L (\F + \FF) \ \int^1_0 {dq \ov q}\ \prod_{I=1}^4  Z_I (\F, \FF; q) \ , 
\  }
where $ c_0 \sim  (2\pi\a')^{-5} V_{10}$, \ 
 $\L(F)$ is the fermionic zero mode factor given  below and  
\eqn\zza{
 Z_I \equiv  \bigg[ { \det ( {\bf 1} - \hat K_I \cdot \hat K_I ) \over
 \det ( {\bf 1}  -  K_I \cdot  K_I)} \bigg]^{1/2} \  }
is the contribution of the non-zero modes. 
The  advantage of  the  GS formalism 
is that the treatment
of non-zero modes of bosons and fermions is  parallel
as both $x$ and $S$ are periodic  functions  of the boundary
 angle $\psi$ (which  is  the euclidean analogue
of the open-string time). 
The bosonic function 
$K_I= K \cdot {\cal F}_I$  ($I=1,2,3,4$) 
 is the product of the derivative of the 
 boundary values   $K_{rs}$
  of the  Green function  of the Laplace operator
on the annulus ($r,s=1,2$)
 \eqn\func{
K_{rs}=  \del_{\psi_r} G_{rs} = -
{ 1 \ov \pi} \sum^\infty_{n=1}   G_{nrs}  \sin n \psi_{rs}
\ ,\ \ \    \ \ \ \psi_{rs} = \psi_r-\psi'_s \ , }
$$    G_n 
\equiv \pmatrix{ A_n &  B_n 
\cr   B_n &  A_n }  \ , 
\ \ \ \ \ \ A_n = { 1 + q^{2n} \ov 1-q^{2n} } \ , \ \ \ \ \ \ 
A^2_n - B^2_n =1  \ ,  $$
with the matrix ${\cal F}_I= \diag (f^{(1)}_I, f^{(2)}_I)$,  where
$f^{(r)}_I$ are the non-vanishing entries in  
$F^{(r)}_{ij}$  taken in the  block-diagonal form\foot{We assume that both $F^{(1)}_{ij}$ and $F^{(2)}_{ij}$
can be simultaneously put in such form.}
$$
F^{(r)}_{ij} = \diag \bigg[ \pmatrix{ 0 & f^{(r)}_1 \cr -f^{(r)}_1  & 0}, 
... , \pmatrix{ 0 & f^{(r)}_4 \cr -f^{(r)}_4  & 0} \bigg]  \ . 
$$
The identity operator  in \zza\ is 
$ {\bf 1}  =  { 1 \ov \pi} \sum^\infty_{n=1}    \cos n \psi_{rs}
\ . $
The fermionic function $\hat K_I$  has a similar definition 
$\hat K_I= K \cdot \hat {\cal F}_I$ 
in terms of $K$ 
(which is also equal to  the  $2\times 2$ matrix of the boundary 
values of the Dirac operator  Green function) 
and 
$\hat {\cal F}_I= \diag (\hat f^{(1)}_I, \hat f^{(2)}_I)$, 
 where $\hat f^{(r)}_I$ are the eigen-values
 of the matrix  $\hat F_{ab}^{(r)}$.

In the case of $\hat F_{ab}$ given by \yyy\  one may use 
 that $i\g^{12}, i\g^{34}, i\g^{56}, i\g^{78}$
are projectors to show  that 
the  non-vanishing elements
$f_I$ and $\hat f_I$ 
in the  block-diagonal forms of 
$F_{ij}$  and $\hat F_{ab}$ 
are related by\foot{This is the same as the 
transformation  in the weight space of $SO(8)$ that  rotates  vectors into spinors, see, e.g., \gsw.
The choice of signs of $\hf_I$ is not important as the determinant 
depends only on their squares (we, in fact, 
invert the sign of $\hat f_1$  as compared to  $\hat F_{12}$
so that   $\prod_{I=1}^4 \hf_I= -\sqrt { \det \hF_{ab}}$).}
\eqn\rela
{ \hf_1 =\ha (  -f_1 + f_2 + f_3  +  f_4) \ , \ \ \ 
\hf_2 = \ha ( f_1 - f_2 +f_3 + f_4) \ , \ \ \  }
$$
\hf_3 = \ha (  f_1 + f_2 -f_3 + f_4) \ , \ \ \ 
\hf_4 = \ha ( f_1 + f_2 +f_3 - f_4) \ ,   $$
i.e.
\eqn\rell{
\hf_I= P_{IJ} f_J \ , \ \ \ \ \ \ 
P_{IJ} \equiv \ha \pmatrix{ -1 & 1 & 1 & 1 \cr
 1 & -1 & 1 & 1 \cr
 1 & 1 & -1 & 1 \cr
 1 & 1 & 1 & -1 \cr}  \ , \ \ \ \   P\inv = P \ , \ \ \ \det P =-1 \ .   }
As  follows from \rell, 
\eqn\tyt{ \sum^4_{I=1}  \hat f_I =   \sum^4_{I=1}  f_I \ , \ \ \ \ \ 
\sum^4_{I=1}  \hat f^2_I =   \sum^4_{I=1}  f^2_I \ . }
Note that it is not necessary to assume that $\hat F_{ab}$ in \yyy\ 
has   block-diagonal form: 
using the projector property of $i\g_{ij}$ one can  compute the fermionic determinant  with the only assumption being that $F_{ij}$ has block-diagonal form.
The result is then  expressed in terms of $f_I$ according to \zza\
where $\hat K_I=K\cdot \diag (\hf^{(1)}, \hf^{(2)})$, 
with $\hat f_I$   defined by \rela.

The  fermionic  zero-mode factor in \zet\ 
is  \refs{\tse,\grw}
\eqn\nuuu{
 \L (F) = -\prod_{I=1}^4 \hf_I= \sqrt { \det \hF_{ab}} 
= - { 1 \ov 3\cdot  2^8} 
 t_8 FFFF   = - {1 \ov 2} 
 \sqrt { \det F_{ij}} + { { 1 \ov 16}} [ F^4 - \fourth (F^2)^2]\ .  }
Since the  matrices $K_I $ and $\hat K_I$
are obtained  from  2$\times $2 matrix  $K$  \func\ 
of the first derivative of the bosonic Green function 
on the annulus with legs on any of the two boundaries
by multiplying it by $\diag({\f, \ff})$
and  $\diag({\hat f^{(1)}_I, \hat f^{(2)}_I})$, 
the ratio of the  functional determinants \zza\ 
reduces to the  infinite product of the ratios of  determinants of $2\times 2$ matrices\foot{There is a misprint in eq.(34) of \tse\ (the equations  that follow (34) are correct): the power of the determinant
should be -1 (as in  eq.(4.9) in \met).}
\eqn\yry{
Z_I =  \bigg[{ \det ( \bI - \hat K_I \cdot \hat K_I ) \over
 \det ( \bI -  K_I \cdot  K_I)} \bigg]^{1/2}
= 
\prod_{n=1}^\infty {\det \hat \Omega_{n,I} \ov \det \Omega_{n,I}} \ , 
}
where
$$\Omega_{n,I} = {\bf 1}  + (K_{n,I})^2\ , \ \ \ \   K_{n,I}
\equiv \pmatrix{ \f A_n & \ff B_n 
\cr  \f B_n &\ff A_n } \ , \ \ \ \ \hat \Omega = \Omega (f^{(r)}_I \to \hat f^{(r)}_I) \ .  
$$
In \tse\ $f^{(2)}_I$ were set equal to zero.
Keeping the non-zero
 background fields at both boundaries  of the annulus 
we get 
\eqn\vtt{
Z_I=  \prod_{n=1}^\infty{ \det \hat \Omega_{n,I} \ov \det \Omega_{n,I}}
= \prod_{n=1}^\infty { (1 - \hat f^{(1)}_I \hat f^{(2)}_I)^2 + (\hat  f^{(1)}_I + \hat f^{(2)}_I)^2 A_n^2 \ov
(1 -  \f \ff)^2 + ( \f+  \ff)^2 A_n^2  }  \ , } 
or, equivalently, 
\eqn\kik{  
Z_I =\bigg( { [1+  (\f)^2][1+  (\ff)^2] \ov
[1+  (\hat f^{(1)}_I)^2][1+ (\hat f^{(2)}_I)^2]} \bigg)^{1/2}\  
\prod_{n=1}^\infty { 1 - 2 q^{2n} \cos 2\pi\tb_I  +  q^{4n} \ov
1 - 2 q^{2n} \cos 2\pi\vp_I  +  q^{4n} }\ , 
}
where  we have used that in  the $\zeta$-function regularisation 
$\prod_{n=1}^\infty  c = c^{-1/2}$.   We defined 
the parameters  $\vp_I$ and $\td \vp_I$  related to $f_I^{(r)}$ and $\hat f^{(r)}_I$   by 
\eqn\dee{
\vp_I = \vp_I^{(1)} + \vp_I^{(2)} \ , \ \ \  \ \ \ 
\tan \pi \vp_I^{(r)} = f_I^{(r)} \ , \ \  }
\eqn\hatt{
\tb_I = \tb_I^{(1)} + \tb_I^{(2)} \ , \ \   \ \ \ 
\tan \pi \tb_I^{(r)} = \hat  f_I^{(r)} \ ,   }
so that 
\eqn\het{
 \tan \pi \vp_I  = { \f+  \ff \ov 1 -  \f \ff }
\ ,  \ \ \ \ \ \ \ 
\cos  2\pi\vp_I = { 1 - \tan^2\pi\vp_I \ov     1  + \tan^2\pi\vp_I }
\ , \ 
}
$$  \sin \pi \vp_I = { \f + \ff \ov \sqrt { [1+  (\f)^2][1+  (\ff)^2]} } \ 
. $$
We reserve 
the notation $\hat \vp_I$ for the linear 
combinations of $\vp_I$ defined as in \rell, 
\eqn\deee{
\hat \vp_I  =  P_{IJ} \vp_J \ , \ \ \ \ \  {\rm  i.e. } \ \  \ 
\hat \vp_1 = \ha ( - \vp_1 + \vp_2 + \vp_3 + \vp_4)\ , ... \ . 
 } 
 Using the expression for  the  Jacobi $\t_1$-function 
\eqn\thet{
\t_1(\vp | i\tau)  
= 
2 q^{1/4} \sin \pi \vp
\prod_{n=1}^\infty  (1- q^{2n})
 (1  - 2 q^{2n} \cos 2\pi\vp +  q^{4n}) \ , 
\ \ \ \ \ q= e^{-\pi \tau}\ ,  }
we finish with 
\eqn\putt{
Z_I= \   { \rf_I  \ov  \hat \rf_I    } \ { \t_1(\tb_I | i\tau)  \ov
\t_1( \vp_I | i\tau) } \ ,  }
where we defined
\eqn\fef{
\rf_I \equiv \f + \ff \ , \ \ \ \ \ \ \ \ \ 
\hat \rf_I \equiv         \hat f^{(1)}_I  + \hat f^{(2)}_I  \  . }
The final result for the partition function \zet\
is thus remarkably simple ($c_1= -\pi c_0$)\foot{Note that the  fermionic zero mode factor $\L(\F + \FF) $ or 
$\prod_{I=1}^4 \hat \rf_I $ effectively  got
`replaced' by its bosonic `analogue'
$ \prod_{I=1}^4 \rf_I $ after we 
expressed the infinite product in terms of the $\t$-functions
(the weak-field expansion of $Z$ is of course 
still proportional to  $\prod_{I=1}^4 \hat \rf_I$).} 
\eqn\zete{
Z= c_1 \int^\infty_0 {d\tau} \ 
\prod_{I=1}^4 \rf_I\  { \t_1(\tb_I | i\tau)  \ov
\t_1( \vp_I | i\tau) } \ . }
\subsec{\bf Green-Schwarz path integral  with supersymmetric 
  boundary condition 
and equivalence with NSR path integral}
It is  easy to show  that if one  replaces
 the  fermionic boundary condition \iop\ by the 
`supersymmetric' one \nnn,\yyyc\  
or, equivalently,  makes the corresponding 
replacement of $\hat F$  in \yyy\  by $\hat F'_{ab}$ such that 
$$\hat M(F)  = { \bI + \hat F' \ov \bI - \hat F'}\ , $$
and  repeats the  above computation of $Z$ 
using \bccs\  and  $\hat F'$ in place of $\hat F$, 
one finishes with the same expression \putt,\zete\
but with 
\eqn\rrr{\td \vp_I \ \ \  \to \ \ \  \hat \vp_I \ , }
where $\hat \vp_I$ is 
defined by \deee, i.e.
\eqn\zetee{
Z= c_1 \int^\infty_0 {d\tau} \ 
\prod_{I=1}^4 \rf_I\  { \t_1(\hb_I | i\tau)  \ov
\t_1( \vp_I | i\tau) } \ . }
To see the reason for the replacement \rrr\
note  first that  the  matrix in the bosonic boundary condition \som\  
  $M= { \bI +  F \ov \bI -  F}$ has  block-diagonal 
form with  four 2$\times $2 entries, 
 $M= M_1 \oplus M_2 \oplus M_3 \oplus M_4$ where 
\eqn\bos{
 M_I=  e^{2\pi  \vp_I  {\cal J}} = \bI \cos 2\pi \vp_I +  {\cal J}  \sin 2\pi \vp_I
= \pmatrix{ { 1 - f^2_I \ov 1 + f^2_I}  &  {  2 f_I  \ov 1 + f^2_I} \cr
 -{  2 f_I  \ov 1 + f^2_I}  &  { 1 - f^2_I \ov 1 + f^2_I}  }  \ , 
\ \ \ \ \   {\cal J} \equiv    \pmatrix{ 0& 1\cr -1 & 0} \ ,  
}
where $ \vp_I ={ 1\ov \pi} {\rm arctan}  f_I$. 
Its  8 eigenvalues are thus
$
e^{\pm 2\pi i  \vp_I }$. 
Since \som\ and \iop\ are related by $F \to \hat F$, \ i.e.\ 
 $f_I \to \hat f_I = P_{IJ} f_J$,  we find 
 that the eigen-values of the `naive' GS fermion 
  rotation matrix $\td M$ in \iop\  are
\eqn\eog{
\td M =\{ e^{\pm 2\pi i \td \vp_I } \} \ .  }
The matrix $\hat M$ in \nnn,\yyyc\
can be represented as\foot{Equivalent form  is \call\
$\hat M = [\sqrt{ \det (1 + F) }]^{-1/2} \exp(2 \hat F) $
where $\hF$ is defined in \yyy\ and the expansion of the 
 exponent is understood in the following sense: $\g_i$ are treated as Grassmann, i.e. they  all anticommute 
and have zero squares.}
 $\hat M = e^{{ 1 \ov 2}\pi  \g^{ij} \vp_{ij}}$, 
where  $\vp_{ij}$  is the matrix whose 8 eigenvalues  
are $\pm i \vp_I$,\  with $\tan \pi \vp_I = f_I$.
Using that $(\g_{2I-1,2I})^2=-\bI, \ [ \g_{2I-1,2I}, \g_{2J-1,2J}]=0$, 
\ $\g_{2I-1,2I} = \g_{2I-1} \g_{2I}$) one  finds \grgu\
\eqn\tyte{
\hat M  =  e^{ \pi \sum^4_{I=1} \vp_I \g_{2I-1,2I}} 
=    \prod_{I=1}^4 ( \bI \cos \pi \vp_I  +   \g_{2I-1,2I} \sin \pi \vp_I)  =
 \prod_{I=1}^4  {1 + 
  \g_{2I-1,2I} f_I \ov \sqrt { 1 + f^2_I } }  \ . }
If  we diagonalise  $\hat M$ (which is just 
a spinor rotation matrix with angles $\vp_I$) 
then  its 8 eigenvalues  will be (cf. \yyy,\rela,\rell)
\eqn\tyr{ \hat M = \{ e^{ \pm 2 \pi i \hat \vp_I} \} \ , \ \ \ \ \ \ \ \ 
  \hat \vp_I= P_{IJ} 
\vp_J \ .  }
Comparing \eog\ and \tyr\
we conclude that replacing \iop\ by \nnn, i.e. 
$\td M \to \hat M$ corresponds to
the  replacement $\td \vp_I \to \hat \vp_I$
in the fermionic GS sector.
The equivalent transformation  $\hat F \to \hat F'$ 
that relates $ \td M= { \bI +  \hat F \ov \bI -  \hat F}$ 
and $ \hat M= { \bI +  \hat F' \ov \bI -  \hat F'}$ 
is determined by  the following non-linear transformation
of the  eigenvalues 
\eqn\repl{
 \hat f_I =P_{IJ} f_J \ \  \to \ \ 
\hat f'_I =  \tan (P_{IJ}\ {\rm arctan} f_J)  =  P_{IJ} f_J  + O(f^2) \ . }
 It would be interesting  to 
find  an independent argument of why  this non-linear redefinition 
 is required.

The equivalent  `proper-time' form of \zetee\
is found by performing  the Jacobi transformation,  
$$
\t_1 ( { i\vp \ov \tt} |{i\ov \tt} ) = i  \sqrt{\tt}\  e^{\pi \vp^2 \ov \tt}\ 
 \t_1(\vp | i\tau)\ ,  $$
and noting that  the exponential  factors
coming from the bosonic and fermionic $\t$-functions
 cancel out because
 of  the  relation between squares of $\vp_I$ and $\hb_I$ 
similar to the one in 
\tyt.\foot{Such relation is not true  for $\tb_I$ and $\vp_I$
 and thus the form of  \zete\  is not `modular-invariant'.}
   We finish with 
\eqn\zettt{
Z= c_1 \int^\infty_0 {dt\ov t^2}\   
\prod_{I=1}^4 \rf_I\    { \t_1(it\hb_I  | it)  \ov
\t_1( it \vp_I| it) } \ , \ \ \ \ \ \ \ \ \ \  t\equiv  {1\ov \tt} \  .  }
In the special case  of 
one non-vanishing  field component this 
 expression was  obtained in \grgu\ 
using  GS boundary state 
approach with the boundary condition \nnn.  

The   expressions \zetee,\zettt\  
 can be put also into   `NSR
form'   by using the 
 Riemann  identity    \kiri
\eqn\jaco{
\prod_{I=1}^4  \t_1 (\hat y_I|it)
=
\ha \prod_{I=1}^4  \t_1 ( y_I|it)  + 
\ha  \sum_{k=2}^4 (-1)^k  \prod_{I=1}^4  \t_k (y_I
|it)\    , }
where four $\hat y_I$ are related to $y_I$ 
as in \rell, i.e.  $\hat y_I = P_{IJ} y_J$. 
Then     from \zetee\ we get 
\eqn\ztep{
Z=\ha  c_1 \int^\infty_0 {d\tau}  
\  \prod_{I=1}^4  { \rf_I \ov
\t_1(  \vp_I| i\tau) }\  
\bigg[  \sum_{k=2}^4  (-1)^k  \prod_{J=1}^4  \t_k ( \vp_J |i\tau)
 +     \prod_{J=1}^4 
\t_1 ( \vp_J|i\tau)   \bigg]  \ , }
and from \zettt\foot{Note that
$\t_2 (-v/z| -1/z) \ov \t_1 (-v/z| -1/z)$=$ -i {\t_4 (v|z) \ov \t_1 (v|z)}$, \ 
$\t_3 (-v/z| -1/z) \ov \t_1 (-v/z| -1/z)$=$ -i {\t_3 (v|z) \ov \t_1 (v|z)}$, 
and 
$\t_2 (-v/z| -1/z) \ov \t_1' (0| -1/z)$=$ { i \ov z} {\t_4 (v|z) \ov \t_1' (0|z)}$, 
\ $\t_3 (-v/z| -1/z) \ov \t_1' (0| -1/z)$=$ {i\ov z}  {\t_3 (v|z) \ov \t_1' (0|z)}$, 
} 
$$
Z= \ 
\ha c_1 \int^\infty_0 {dt\ov t^2}  \  
\prod_{I=1}^4  { \rf_I \ov
\t_1( it \vp_I| it) }  \bigg[  \prod_{J=1}^4  \t_2 (it \vp_J |it)
- \prod_{J=1}^4  \t_3 (it \vp_J |it) + 
 \prod_{J=1}^4  \t_4  (it \vp_J |it)\bigg] $$
\eqn\zte{
+ \  \ha c_1 \int^\infty_0 {dt\ov t^2}  \ 
\prod_{I=1}^4   \rf_I  \ . }
Apart from the last term, this is, indeed, the result which one finds 
by doing a similar calculation in the NSR approach. 

The last term  (i.e. $\sim \sqrt{\det \rF_{ij}}$ 
which does not have  a  $D=10$  Lorentz-covariant 
extension)  
does not   appear in the standard
D9-brane  NSR partition 
function in magnetic background 
(the periodic sector contribution  
vanishes because of the remaining fermionic zero modes $\psi_0$,$\psi_9$)
but is present in the closely related
  D8-brane  expression \refs{\lif,\pierre,\berli,\bill}, 
obtained by  assuming the  Dirichlet boundary  condition along 
 the 9-th direction. 
The reason  why it appears in the   GS  approach
is that because of the choice of the l.c. gauge 
here  one  treats 
the   $0,9$ directions as the Dirichlet ones \refs{\grr,\grgu}, i.e.  
 the above expression \zettt\ effectively 
corresponds to  D8-instanton, with time being  one of the two 
orthogonal directions. 

In NSR approach  the  fermionic
terms in  the action 
in \gen,\ope\  are  replaced by ($\m,\n=0,1,..,8,9$)
\eqn\ses{ {i \ov 8 \pi}  \int  d^2 \sigma ( \psi^\m_R \del_+ \psi^\m_R + \psi^\m_L \del_- \psi^\m_L)
 - {i\ov 2} \a'   \int dt\  F_{\m\n} \psi^\m \psi^\n \ ,    }
with the boundary conditions 
$\psi^\m_R= \psi^\m_L= \psi^\m $ at $\sigma =0$ and 
$\psi^\m_R= \mp \psi^\m_L= \psi^\m $ at $\sigma =\pi$. 
The  calculation of the
fermionic determinants is analogous to the one  discussed above
but now  we are   to sum over the contributions of different sectors. 
Since here  we have  the same  $F_{\m\n}$ matrix  appearing 
in the bosonic and fermionic determinants, the arguments
of the $\t$-functions are also  the same. For $F_{\m\n}$ having 
only magnetic $F_{ij}$ ($i,j=1,...,8$) components 
one finds that the final expression for the 
 partition function is given by
\zte\ (without the last term).

The NSR approach allows, in principle, to obtain 
the    expression for $Z$
for the general $D=10$  choice of  (euclidean) $F_{\m\n}$ having 
 all five `eigenvalues' $f_0,f_1,...,f_4$
being non-vanishing ($f_0= F_{09} = iE$  is  the electric field 
component).\foot{In the case of Minkowski signature 
there is another irreducible form of constant $F_{\m\n}$ 
which is  the direct superposition of  the block-diagonal magnetic 
field strength  $F_{pq}$ \   ($p,q=5,...,9$)
and the 4-dimensional `plane-wave'  field $F_{\alpha\beta}$ \
 ($\a,\beta=0,1,2,3$) with  
orthogonal electric and magnetic components,  e.g.,  $F_{+3}=0$, i.e.  
 $F_{03}=F_{13}=\const$. The open bosonic string spectrum in such 
 background was studied in \kosh. 
 Since the  non-trivial 4-dimensional part of this background
 preserves supersymmetry, the  corresponding superstring partion
 function vanishes. The bosonic string partition function is also trivial 
 since all Lorentz-invariant scalars  vanish when evaluated on the 
  plane-wave background 
 $F_{\a\b}$.}
The natural generalization of \zte\ to the case of $f^{(1)}_0, f^{(2)}_0\not=0$ is 
\eqn\zeg{      Z
= \ 
\ha   c_1 \int^\infty_0 {d\tau}  \ 
    \prod_{M=0}^4 { \rf_M
         \ov   \t_1 ( \vp_M|i\tau) }\  \bigg[
       \sum^4_{k=2} (-1)^k   {  \t'_1 (0 |i\tau)  \ov \t_k ( 0 |i\tau) } 
  \prod_{N=0}^4 \t_k ( \vp_N |i\tau)   +       \prod_{N=0}^4 \t_1 ( \vp_N |i\tau) \bigg]  } 
or, equivalently, after the modular transformation  ($ t=1/\tt$) 
$$      Z
= 
\ha    c_1 \int^\infty_0 {dt\ov t^2}  \ 
    \prod_{M=0}^4 { \rf_M
         \ov   \t_1 ( it  \vp_M|it) }\  \bigg[\ 
      it  \sum^4_{k=2} (-1)^k   {  \t'_1 (0 |it)  \ov \t_k ( 0 |it) } 
  \prod_{N=0}^4 \t_k ( it\vp_N |it)  
  +  \prod_{N=0}^4 \t_1 ( it\vp_N |it) \bigg]  $$ 
$$
= \ 
\ha i  c_1 \int^\infty_0 {dt \ov t}\   \prod_{M=0}^4  \rf_M \ 
  \bigg[\ 
  {  \t'_1 (0 |it) 
         \ov \t_2 ( 0 |it) } 
\prod_{N=0}^4  { \t_2 (it \vp_N |it) \ov
                \t_1 (it \vp_N |it) }
- \  {  \t'_1 ( 0 |it) 
         \ov \t_3 (0  |it) } 
\prod_{N=0}^4  { \t_3 (it \vp_N |it) \ov
                \t_1 (it \vp_N |it) }
$$
\eqn\zteg{ 
+ \ {  \t'_1 ( 0|it) 
         \ov \t_4 ( 0 |it) } 
\prod_{N=0}^4  { \t_4 (it \vp_N |it) \ov
                \t_1 (it \vp_N |it) }\bigg]
+\ \ha   c_1  \int^\infty_0 {dt\ov t^2}\    \prod_{M=0}^4  \rf_M     \ .   }
The $D=10$  expressions \zeg\ and \zteg\ 
  which are  symmetric in all  
five field strength eigenvalues  reduce to  \ztep\ and  \zte\
 in the limit 
$f_0^{(1,2)} \to 0$,\  \ $\vp_0\equiv {1\ov \pi } (\arctan   f_0^{(1)} + \arctan   f_0^{(2)}) \to 0$.  
Some of the  properties 
of the partition function 
\zteg\ will be studied in sections  2.3  and 2.4 below.
In particular, it has the correct $D=10$  SYM ($\a'\to 0$)
 limit.

The  special  case  of the superstring partition function \zte,\zteg\
 when only  
one   (electric, $f_0 =i E$) gauge field  
component is non-vanishing  was first obtained  in \bacp\ 
(instead of computing the
 determinants directly with field-independent boundary conditions,
  the authors of \bacp\ followed the equivalent 
  procedure of solving the string equations with 
field-dependent boundary conditions   \refs{\callan,\call,\nester}, i.e.  \som\ and 
$\psi^\m_R-  \psi^\m_L =  F_{\m\n} (\psi^\m_R+  \psi^\m_L)$ at $\sigma=0$, 
and
$\psi^\m_R \mp  \psi^\m_L =  F_{\m\n} (\psi^\m_R\pm   \psi^\m_L)$ at $\sigma=\pi$
and   explicitly  determining 
the corresponding  string spectrum).
The resulting expression was applied in D-brane context in \bac\ 
and  was extended   to the case of  four or less  
 non-vanishing  eigenvalues of 
 $F_{\m\n}$   in  \refs{\lif,\pierre}
(ref. \pierre\ gave also 
partial $f_0 << f_I$  result  for  the general five-eigenvalue  case).
 It is easy to check that 
the expressions in \refs{\bacp,\bac,\lif,\pierre}
are indeed the appropriate special cases of \zte.

\subsec{\bf D-brane applications  and some properties of $Z$ }

The partition function \zettt,\ztep,\zteg\ has several applications.
When the boundary background fields  are equal  $F^{(1)}=F^{(2)}=F$ it 
may be interpreted as a  one-loop  contribution 
to the tension of D9-brane  with some background field 
distribution $F$.  Similar expressions are  found for other   Dp-branes
by setting some field components to zero 
and adding extra factor $s^{(9-p)/2}$ 
in the integration measure
(reflecting the Dirichlet nature  of $9-p$ transverse directions). 
  If instead  one introduces the 
open string mass factor $\exp( -M^2 t),  $ with  $M^2 = {r^2\ov 2\pi \a'}$ 
representing  the 
separation  $r$  between parallel D-branes \pol,\  
one finds  the potential between two Dp-branes with  generic backgrounds 
$F^{(1)}$ and $F^{(2)}$ on each brane
(keeping   the transverse electric field component non-vanishing
allows  also to describe the potential between moving branes \bac).
Various other cases are obtained as special ones 
by sending  some of the field  components
$f^{(r)}_I$ to zero or infinity.

For example, for  two parallel D8-branes one finds
\eqn\ptt{
Z= c_2 \int^\infty_0 {dt\ov t^{3/2} }\  e^{-{r^2 \ov 2\pi \a'}t} \   
\prod_{I=1}^4 \rf_I\    { \t_1(it \hat\vp_I | it)  \ov
\t_1( it \vp_I| it) }  \ ,   }
where $c_2 ={   2 V_9\ov (2\pi^2 \a')^{1/2} (2\pi\a')^4 }$ and 
 $\rf_I, \vp_I, \hat \vp_I$ are defined in \fef,\dee,\deee.
If $f^{(2)}_I=0$  while $f^{(1)}_I$ is generic 
we get the  potential between `pure' D8-brane and  a non-marginal 
bound state 8+6+...+0 of branes. 
It is clear from \gen,\ope\  or from the form of the boundary
conditions \som, that sending the boundary components 
 $f^{(2)}_I$  to infinity is equivalent to changing from Neumann to Dirichlet
conditions  in   the 
 two directions $x^{2I-1},x^{2I}$  on the second brane. This then 
gives the potential between D8-brane and D6-brane (or 
between D8-brane and D0-brane if $f^{(2)}_I\to \infty$ for
 all $I=1,2,3,4$).\foot{For 
    $f^{(2)}_I=0$  one has  $\vp_I= \vp_I^{(1)}$,  while 
      for $f^{(2)}_I \to \infty $ \het\ implies that 
$\tan \pi \vp = - 1/f^{(1)}_I$, i.e. $ \vp_I = \vp_I^{(1)} -\ha $, 
so that  one can use 
$\t_1( \vp-\ha|i\tau ) = -\t_2 (\vp|i\tau) , \ 
\t_3( \vp -\ha|i\tau ) = \t_4 (\vp|i\tau)$.}
The limit 
$f^{(r)}_I\to \infty $  on both boundaries 
corresponds to performing T-duality in the ${2I-1},{2I}$ 
directions and thus gives potential between two D6-branes (or D7-branes if 
the starting point is D9-brane expression), etc. 

  The last zero-mode term in \zteg\ 
proportional to  $\prod_{M=0}^4  \rf_M  = \sqrt{\det{ \rF_{\m\n}}}$ 
reduces to  the last term in (euclidean) D7-brane  analogue of \zte\
if one sends $\rf_1$ ($\rf_0$) to infinity which is equivalent to 
performing  T-duality transformation
along  the two orthogonal  directions.\foot{In this limit 
one  should fix the product of the field component
 with the volume in the orthogonal  directions
so that the finite  coefficient in front of the (euclidean)  D7-brane 
analogue of \zte\ is 
proportional to $V_8$. Analogous  considerations can be used 
to relate D9-brane action to  D8-brane one.
Similar term  (which is the contribution of $R (-1)^F$
sector) 
  is present in the 
 interaction potential  between D0-brane and D8-brane
 \refs{\lif,\pierre,\berli,\bill}
and the related case of interaction between D4-branes at angles \sheih.}
The analogs of the last term in \zteg\
(which is  absent in the  SYM theory limit) 
appear also in the case of lower-dimensional  D-branes
and   may be  
related to the  top  `Chern-Simons' terms 
which multiply the RR fields  in D-brane actions
\refs{\call,\li}.


The superstring partition function  \zet,\zettt,\zteg\ 
vanishes not only in the zero-field limit  but also  
in the `neutral string' $\rF\equiv F^{(1)} + F^{(2)}=0$  limit
($\rf_I=0$ implies $\vp_I=0$ and $\hat \vp_I=0$, see \dee,\deee).\foot{
This is in contrast to  what happens in the  neutral 
bosonic  string case where 
$Z (F)  = \det (\d_{ij} + F_{ij}) Z(0)$ \callan. The one-loop $Z$
 in the open bosonic string  has non-trivial dependence 
 on the fields if $F^{(1)} \not=- F^{(2)}$. \  $Z$ with  
 $F^{(1)} = F^{(2)}$ was  found  
 in \ft;   the case of  $\FF=0$ 
and its  relation  to   YM  effective action in the $\a'\to 0$ limit
was studied in \met\ (some related computations  of bosonic 
string partition function in external electromagnetic field 
and at finite temperature 
appeared in \refs{\odn,\odd}).}
In   D-brane context this corresponds to the vanishing of the 
potential between two parallel Dp-branes with the same field
backgrounds (i.e. the same extra RR charge distributions) or the 
vanishing of  potential  between
two parallel Dp-branes moving with 
the same velocities (equal to electric fields after T-duality).

As follows  from \zet,\nuuu,
the `magnetic' partition function \zettt\ 
vanishes  when any  of the parameters  $\hat \vp_I $
is zero,   i.e.  for the gauge field  backgrounds which have 
 residual  supersymmetry  from string-theory point of view, 
cf. \nnn,\tyr\ 
\refs{\grgu,\call} (see also, e.g.,  \refs{\dougl,\sheih}). 
 Note  that, in general, the condition  that some 
 $\hat \vp_I = {1\ov \pi} P_{IJ} 
(\arctan f^{(1)}_J + \arctan f^{(2)}_J)= 0$ 
is not equivalent to the vanishing 
of some of  $\hat \rf_I=P_{IJ} \rf_J$, i.e.  to 
the   SYM theory supersymmetry condition
 (when  $\hat \rf_I=0$ the corresponding  matrix 
$\hat \rF_{ab} = \fourth \g^{ij} \rF_{ij}$ 
has zero modes).\foot{The two conditions are equivalent,  however, 
in  the most interesting 
cases, e.g., 
 for the 4d instanton configuration
($\rf_1=\pm \rf_2, \ \rf_3=\rf_4=0$  implies $\vp_1=\vp_2, \ 
  \vp_3=\vp_4,$  and thus $\hat \vp_1=\hat \vp_2=0$).} 
 For $F^{(1)}=F^{(2)}$
this  implies, in particular,  the vanishing of the 
one-loop contribution to 
the tension of a supersymmetric bound state of Dp-branes,
  while  for  generic 
$F^{(1)}$ and $F^{(2)}$ this implies the vanishing of the potential 
for a supersymmetric  configuration of Dp-branes.


Let us now  discuss  the  $t\to 0$ (open string channel) and $t \to \infty$ 
(closed string channel) 
behaviour  of the integral in  $Z$ \zteg\ for arbitrary background fields.
 Let us first 
consider the `magnetic' case when  $Z$ is given by  \zettt,\zte.
 In the  $t\to 0$ ($\tt\to \infty$) region 
 it is clear from the representation 
\zetee\ and \thet\  that the integral in \zettt\ reduces to 
\eqn\what{
Z\ \to \  \int_{t\to 0} {dt\ov t^2} \  
\prod_{I=1}^4  \rf_I\  { \sin \pi  \hat \vp_I \ov  \sin \pi \vp_I   }
  \  .   }
The weak-field expansion\foot{In general, the 
  the weak-field 
limit $Z$  \zetee\
 is  proportional to $\prod_{I=1}^4 \hat \rf_I$, i.e. to 
 $ \sqrt{\det \hat \rF_{ab}}$.
 }
 of this expression starts with 
$\prod_{I=1}^4 \hat  \rf_I$, 
i.e.   $Z$  contains  
  $O(F^4)$  (quadratic)  UV  divergence  (cf. \nuuu)   which in type I string  theory 
 is canceled  as in \green\  
 after adding the M\" obius strip  contribution \tse. 
In Dp-brane context, the measure gets extra factor of $s^{1/2}$ for each of the 
$9-p$ 
Dirichlet  directions    so  that the integral becomes 
convergent  at $t\to 0$  for $p < 7$.\foot{Note that the fact that the 
leading  $F^4/M^{7-p}$  term 
(in the integral with factor $e^{-M^2 t}$ included)
 originates
from the zero-mode factor explains  its  `universality', i.e. its 
trivial scale-factor  dependence on $\a'$ or $r$ (cf.   \refs{\dkps,\baki}).}
 
For  large $t$  \zettt\ and \thet\  lead to a similar 
expression but now with $\vp_I\to it \vp_I$:
\eqn\simi{
Z\ \to \  \int^{t\to \infty} {dt\ov t^2} \  
\prod_{I=1}^4  \rf_I\  { \sinh \pi t \hat  \vp_I \ov  \sinh \pi t \vp_I
   }   \  .   }
This is convergent when, e.g.,
 all $f^{(r)}_I$ are such that  all $\vp_I$ and $\hat \vp_I$ 
have the same sign: the  divergence cancels out 
because  of $\sum^4_{I=1} (\hat \vp_I - \vp_I)=0$ (cf. \tyt). 
 If only one  (e.g., the  first) component  of $\rf_I$ and thus of 
$\vp_I$  is non-vanishing,  
one finds that the integral is divergent 
at large $t$. Introducing the  IR cutoff factor
$e^{-M^2 t}$  we get 
\eqn\simil{
Z\  \ \to \ \   \rf_1  \int^{t\to \infty} {dt\ov t^5} \  e^{-M^2 t}\ 
   { \sinh^4 {1\ov 2} \pi t\vp_1  \ov  \sinh \pi t \vp_1 
   }  \ \sim  \ \ \rf_1
 \int^{t\to \infty} {dt\ov t^5} \  e^{-M^2 t} \ 
   e^{  \pi   |\vp_1 | s } + ...  \     . }
The integral is convergent  if
$ M^2 > \pi  |\vp_1 |$
(similar remark 
was made in D-brane context in \grgu). 
For $M=0$ the  resulting divergence 
 is the string analogue of the well-known  
IR instability \nilol\ 
of the YM theory  in magnetic backgrounds
 (which remains also  in the  super YM theory \frao).

In general, starting with the full $D=10$  partition function
\zteg\ we find in the $t\to \infty$ limit\foot{We use that for $t\to \infty$ 
\ \ \ \ 
$\t_1 (iz |it) \to 2i  e^{-{ \pi t \ov 4}}\sinh \pi z,  \ \
 \t_2 (iz |it) \to 2  e^{-{ \pi t \ov 4}}\cosh \pi z,  $ \  \ \ 
 $ 
\t_3 (iz |it) \to 1+ 2  e^{-{ \pi t}}\cosh 2 \pi z,   \ \ 
\t_4 (iz |it) \to 1  -2  e^{-{ \pi t}}\cosh 2\pi z.  $}
$$Z \to  \ 
\ha   c_1 \int^{t\to \infty }
 {dt \ov t}  \ 
  {  
e^{- {\pi t\ov 4}}  \prod_{M=0}^4 \rf_M 
         \ov  
 16 \prod_{M=0}^4 e^{- {\pi t\ov 4} } \sinh \pi t \vp_M
} 
\bigg[\ 
 16 e^{ {\pi t\ov 4}} \prod_{N=0}^4   e^{- {\pi t\ov 4} }\cosh  \pi t \vp_N  
    $$
$$
-  \   {\prod_{N=0}^4 (1 +  2 e^{- {\pi t}} \cosh 2 \pi t \vp_N)\ov
  1 +  2 e^{- {\pi t}} }
\ + \  {\prod_{N=0}^4  (1 -  2 e^{- {\pi t}} \cosh 2  \pi t \vp_N)\ov
  1 -  2 e^{- {\pi t}}
 } \bigg]
+  ... $$ 
\eqn\tre{
\to
\ 
{1\ov 8}    c_1 \int^{t\to \infty }
 {dt \ov t}  \ 
  {  
  \prod_{M=0}^4 \rf_M 
         \ov  
  \prod_{M=0}^4  \sinh \pi t \vp_M
} 
\bigg[\ 
 4  \prod_{N=0}^4  \cosh \pi t \vp_N  
-     \sum _{N=0}^4  \cosh 2 \pi t \vp_N   + 1   \bigg] + ...  \ ,   }
were we kept only the leading term in each $\t$-function factor
and dropped the last term in \zteg\ which is convergent at $t\to \infty$.  
In   the euclidean case with 
all $\vp_M$  being  real   we find that the condition 
of  convergence at 
$t\to \infty$ 
is determined by the term 
$ [\prod_{M=0}^4  \sinh  \pi t \vp_M]\inv  \sum _{N=0}^4  \cosh 2 \pi t \vp_N$.
Assuming   that all  $0<\vp_M < \ha $,  the integral is thus convergent if 
$ \vp_N - \sum_{M\not=N} \vp_M  < 0$  for each $N$.

If the electric fields are real, i.e. if  
 $f^{(r)}_0 = i E^{(r) } $ and thus $\vp_0= i \vp_e$ are imaginary, 
we get the  infinite number of poles coming  from  the bosonic 
$1\ov \sin \vp_e t$
factor. This is the open (super)string analogue of
 the well-known Schwinger  
 pair-creation instability in  electric field 
  \refs{\burg,\bacp}. 
  Since in this case 
  $  \tanh \pi \vp^{(r)}_e  =    E^{(r)}$, 
  there is the  
  restriction $ E^{(r)} \leq 1$ 
  (at the critical value of the electric field 
  $\vp^{(r)}_E$ goes to infinity and the  production rate diverges
  \bacp).

\subsec{\bf Super Yang-Mills  theory limit  }
Let us  now show that 
 the $\a'\to 0$ limit of 
 the string theory partition function \zete,\zetee,\zteg\  
  reproduces  the $D=10$ super Yang-Mills one-loop effective action 
in  constant abelian background.
As discussed in \refs{\met}, the  $D=10$ field-theory limit
 corresponds to 
$\a' \to 0$ for fixed UV cutoff at $t=0$,\foot{In general, 
one should also fix 
$g_{YM}^2 = g_s \a'^{(D-4)/2}$ in the limit.
 There is no cutoff dependence 
in $D< 8$ \bgs.} i.e.  to 
$t\to \infty$ (or  $\tt\to 0$). 
We shall  restore the factors $2\pi \a'$ 
in $F_{ij}$, i.e. $f_M \to 2\pi \a' f_M$,  and
  define ${\rm t}  = 2\pi \a' t $ which has   dimension (length)$^2$
   so that 
$ 2\pi \a' f_M t = f_M \s$ 
is  dimensionless.  The parameter 
 $\s$  is the (open string) field theory proper-time parameter, 
$ {\ep^2} \leq \s < \infty, $  where $ \ep$  is the field-theory UV cutoff.
Taking $\a'\to 0$
for fixed $\s$ and $\ep$  
we  find  that (see \dee) \ 
$ \vp_M \to 2 \a' \rf_M$, \  
$ \hat \vp_M \to 2 \a' \hat \rf_M$, 
 so that  the  limiting value of  $Z$ 
  depends only on the sum of the two 
boundary fields $f^{(1)}_M + f^{(2)}_M$=$\rf_M $.

Since in this limit $t= {\s \ov 2\pi \a'} \to \infty$
while $  2\pi \a' f_M t = f_M \s \to \pi \vp_M \s$  is fixed, 
the resulting limiting form of $Z$ \zteg\ is essentially given by 
\tre\ (with $\pi  \vp_M \to  \rf_M$ and 
extra overall factor of $(2\pi \a')^{5}$ coming 
out of $\prod_{M=0}^4 \rf_M$).  This is 
to be compared with the $D=10$ SYM 
analogue of the one-loop background field  effective action. 
In the case of the $SU(2)$ theory with (euclidean) 
 $U(1)$  $D=10$ background
$F_{\m\n} = 
\ha {\sigma_3} {\rm F}_{\m\n}$,  where ${\rm F}_{\m\n}$ 
has block-diagonal form with 5 non-zero entries 
$\rf_M$,    one finds for the one-loop SYM effective action 
  \refs{\chets,\zarem}\foot{Similar expressions  in $D < 10$ SYM theories 
are found after one corrects  the power
 of $\s$ in the measure ($1\ov  \s$ $\to$ $1\ov  \s^{1 + {1\ov 2}D}$) 
 to take into account 
the effect of dimensional reduction (or, equivalently, sets
the components  $\rf_M$ corresponding to `extra' $D-10$ dimensions
to zero).}
\eqn\ter{ \Gamma (F)    =  - { 2 V_{10} \ov  (4\pi)^{5} }
\int_{{\ep^2}}^{\infty }{d\s \ov \s }\  \prod^{4}_{M=0}{ \rf_M  \ov \sinh \rf_M\s}\
 \bigg(
 \sum_{N=0}^{4}  \cosh 2\rf_N\s \  - \  1 -  \  4\prod^{4}_{N=0} \cosh\rf_N\s\  
 \bigg)\ . }
This is in   precise  agreement  with the expression that  follows from 
\zteg,\tre\  with the  normalisation of the string-theory result  fixed as  
\eqn\coff{
c_1 = { V_{10}\ov 2  (2\pi)^{5} (2\pi \a')^{5}} \ . }
Note that 
the last  $\sqrt{ \det { \rF_{\m\n}}}$ term  in \zteg\
has no counterpart in the SYM theory:  it disappears  
 in the $\a'\to 0$ limit (it gets one extra power of $\a'$ after $t \to {\s \ov 2\pi \a'}$
and $\rf_M \to 2\pi \a' \rf_M$).

Let us  consider explicitly  the purely magnetic case ($\rf_0=0$).
Both   expressions \zete\ and \zetee\  for the `magnetic'   GS 
partition function  have the same   
field-theory   limit
(both $\td \vp_I$ \hatt\  and  $ \hat \vp_I$ \deee\  are equal  to 
$2 \a' \hat \rf_I$ in the $\a'\to 0$ limit). 
Since
 $\t_1({i \rf_I \s\ov \pi}  |{ i\s \ov 2\pi\a'})_{\a'\to 0}  \ \ 
 \to  \ \ 2i \ e^{- {\s \ov 8\a' } }
\sinh  \rf_I\s \ ,  $
we  get that in the limit  $\a'\to 0$  $Z$ becomes 
\eqn\zetty{
Z  = c_0 \int^\infty_{\ep^2} {d\s\ov \s^2} \  
\prod_{I=1}^4  \rf_I\ \bigg( \prod^4_{J=1} { \sinh  \hat \rf_J\s   \ov  \sinh  \rf_J\s } - {1\ov 2} \bigg) 
  \ , 
\ \ \ \ \ \   c_0 =  (2\pi \a')^5 c_1 \ .    }
The term  $-\ha$  in the bracket is  introduced  to  subtract  the 
last $\sqrt {\det{ \rF_{ij}}}$  term in \zte\ 
which  has a non-vanishing  $\a'\to 0$ limit 
but is not reproduced by  the  SYM theory
(as was mentioned above, it is absent in the $\rf_0 \to 0$ limit
of the covariant D9-brane NSR  expression \zteg\ 
but appears in  related euclidean   D8-brane expression). 
Like \ter, 
this integral has the  well-known   UV divergence of  SYM theory
(quadratic in $D=10$ and logarithmic in $D=8$)
  which is proportional to  (cf. \nuuu) \ 
$  F^4 - \four (F^2)^2$\   
  \refs{\frao,\met}.  
Eq. \zetty\   is indeed equivalent  to the $\rf_0 =0$ limit of 
\ter\ 
\eqn\tert
{
 \Gamma (F)    =  - { 2 V_{10} \ov  (4\pi)^{5} }
\int_{{\ep^2}}^{\infty } {d\s \ov \s^2 }\ 
 \prod^{4}_{I=1} { \rf_I \ov \sinh \rf_I \s}\
 \bigg(\sum_{J=1}^{4}\cosh 2\rf_J\s - 4\prod^{4}_{J=1} \cosh \rf_J\s \bigg)
\ ,   }
since,  as follows from \rela, 
\eqn\uyy{
8 \prod_{I=1}^4  \sinh  \hat \rf_I\s - 4  \prod_{I=1}^4 \sinh   \rf_I\s =
  - \sum_{I=1}^{4}\cosh 2\rf_I\s  +  4\prod^{4}_{I=1} \cosh \rf_I\s  \ . 
    }
We shall later use a similar identity (obtained from \uyy\ by  
$f_k \to f_k + {i\pi\ov 2}$) 
\eqn\uyyu{
8 \prod_{I=1}^4  \cosh  \hat \rf_I\s - 4  \prod_{I=1}^4 \sinh   \rf_I\s =
   \sum_{I=1}^{4}\cosh 2\rf_I\s  +  4\prod^{4}_{I=1} \cosh \rf_I\s  \ . 
    }
 Eq. \zetty\    gives a  useful  expression
for  the 1-loop effective action in $D=10$ SYM theory
in purely magnetic (or  `8-dimensional')  background. 
In particular, it is clear from it   that the 
effective action   becomes simply proportional to 
$\prod^4_{I=1} \rf_I$ 
 for all supersymmetric abelian  gauge field   backgrounds
 for which $\hat F_{ab}= \four \g^{ij}_{ab}  F_{ij}$ has
zero modes, i.e.   for which 
 some of $\hat f_I$ vanish 
(in particular, \zetty\ 
 vanishes  for the  $D=4$ instanton background, $f_1=\pm f_2$, 
but, in contrast to \zettt,\zte,  
 is non-vanishing 
for  its `$D=8$ generalisation',  
$f_1=\pm f_2, \ f_3=\mp f_4$).

The special case of \zetty\ with $\rf_3=\rf_4=0, \ \rf_2 \to \rf_0$
(i.e. $\hat \rf_1=-\hat \rf_2 = \ha (\rf_0 - \rf_1), \
\hat \rf_3=\hat \rf_4 = \ha (\rf_0 + \rf_1)$)
is directly related to 
  the effective action of the four-dimensional 
$N=4$ \  $SU(2)$ SYM 
theory in  constant $U(1)$ background \frao\ 
\eqn\tekr{ \Gamma(F)    =  -{  V_{4} \ov  \pi^{2} }
\int_{0}^{\infty }{d\s \ov \s }\  \rf_0 \rf_1\ 
{  \sinh^2 {\rf_0 - \rf_1 \ov 2}\s  \  \ \sinh^2 {\rf_0 +\rf_1 \ov 2}\s \ov
 \sinh \rf_0 \s\ \  \sinh \rf_1 \s } \ . }
Here $\rf_0$  stands for the euclidean  analogue 
of the $D=4$ electric field component. This integral 
is convergent at $\s \to 0$ but has the same magnetic 
IR instability at $\s \to \infty$ as in YM theory.



\newsec{FINITE TEMPERATURE CASE}
\subsec{\bf Free energy  of Super Yang-Mills  theory}
 It is useful  first  to  recall  that in a  $D=p+1$ dimensional 
field  theory  with 
 bosonic  and fermionic degrees of freedom  
with   mass operators $\hat M_B$ and $\hat M_F$
(which may depend on a background field)
the proper-time representation for the 
free energy $\cF$ 
has the form
\eqn\field{
Z = \b \cF =  - {V_{p}\b \ov 2(4\pi)^{(p+1)/2}}
 \int^\infty_0 { d\s \ov \s^{p+3 \ov 2}}\  \bigg[
 \t_3 ( 0| {i\b^2 \ov 4 \pi \s}) 
  \     \Tr\  e^{- \s \hat M^2_B} 
-
 \t_4 ( 0| {i\b^2 \ov 4 \pi \s}) 
  \     \Tr\  e^{- \s \hat M^2_F}  \bigg]
\ .  }
The inverse temperature $\b$ is the period of the euclidean time direction.
Using that 
\eqn\ide{
\t_3 (0| iz) \pm  \t_4 (0|iz) =  \sum^\infty_{n=-\infty} [1 \pm (-1)^n]
 e^{- \pi z n^2} =  2 \t_{3,2} (0| 4iz) \ ,  }
we can represent \field\ in the form
\eqn\ield{
Z =   - {V_{p}\b \ov 2(4\pi)^{(p+1)/2}}
 \int^\infty_0 { d\s \ov \s^{p+3 \ov 2}}\  \bigg[
 \t_3 ( 0| {i\b^2 \ov  \pi \s}) 
  \    ( \Tr\  e^{- \s \hat M^2_B} - \Tr\  e^{- \s \hat M^2_F})
}
$$
\ \ \ \ \ \ \ \ \   +  \ 
 \t_2 ( 0| {i\b^2 \ov  \pi \s}) 
  \      ( \Tr\  e^{- \s \hat M^2_B} +  \Tr\  e^{- \s \hat M^2_F})
 \bigg]
\ ,   $$
or, equivalently, as 
\eqn\ield{
Z =   - {V_{p} \ov 4(4\pi)^{p/2}}
 \int^\infty_0 { d\s \ov \s^{p+2 \ov 2}}\  \bigg[
 \t_3 ( 0| {i \pi \s \ov \b^2}) 
  \    ( \Tr\  e^{- \s \hat M^2_B} - \Tr\  e^{- \s \hat M^2_F})
}
$$
\ \ \ \ \ \ \ \ \   +  \ 
 \t_4 ( 0| {i \pi \s \ov \b^2}) 
  \      ( \Tr\  e^{- \s \hat M^2_B} +  \Tr\  e^{- \s \hat M^2_F})
 \bigg]
\ .   $$
In the  zero-temperature limit $\b\to \infty$  only the first  term in \ield\
survives ($ \t_2 ( 0| {i\b^2 \ov  \pi \s})  \to 0 $, 
 $\t_{3,4}  ( 0| {i\b^2 \ov  \pi \s})\to 1 $), i.e.
  \ield\
 reduces to the standard  integral of 
$\Tr\  e^{- \s \hat M^2_B} -  \Tr\  e^{- \s \hat M^2_F}$.
In the special case of free supersymmetric theory with equal 
masses  of bosons and fermions $\hat M_B =\hat M_F= \hat M$  this becomes
\eqn\eld{
Z =   - {V_{p}\b \ov (4\pi)^{(p+1)/2}}
 \int^\infty_0 { d\s \ov \s^{p+3 \ov 2}}\  \t_2 ( 0| {i\b^2 \ov  \pi \s}) 
  \    \Tr\  e^{- \s \hat M^2}  \ . 
}
If the  fields are   massless
 with  total  $\cN$ of   bosonic degrees of freedom 
($\cN=8N^2$  for the  $U(N)$ SYM theory) 
\eqn\zzee{
Z(\b)  = -\cN \k_p V_p \b^{-p} \ , \ \ \ \ \ \ \ \ 
\k_p = [1 + (1 - 2^{-p}) ] (2\pi)^{-p} \om_{p-1}(p-1)!\ \zeta(p+1) 
 \ ,  }
where $\om_{p-1} ={ 2 \pi^{p\ov 2}\ov \Gamma({p\ov 2})}$.  

 The one-loop  $SU(2)$  SYM
effective action   \ter,\tert\ is given by the  sum 
of contributions of  bosonic and fermionic determinants.
Observing that the fermionic contribution  in \tert,\ter\
is represented by the term proportional  to $\prod_{J} \cosh \rf_J\s, $
we find the  following finite-temperature generalisation 
of  \tert, i.e. the one-loop free energy  of the 
$D=10$  SYM theory in a magnetic background (cf. \field) 
$$
Z (\b,F)  = \b \cF(\b,F)
=  -  { 2 V_{9} \b \ov  (4\pi)^{5} }
\int_{{\ep^2}}^{\infty } {d\s \ov \s^2 }\ 
 \prod^{4}_{I=1} { \rf_I \ov \sinh \rf_I \s}\
$$
   \eqn\tykl{\times\   \bigg[
 \t_3 ( 0| {i\b^2 \ov 4 \pi \s})   \sum_{J=1}^{4}\cosh 2\rf_J\s - 
 \t_4 ( 0| {i\b^2 \ov 4 \pi \s})\   4\prod^{4}_{J=1} \cosh \rf_J\s 
 \bigg]
\  .   }
Using \uyy,\uyyu,\ide\ we can  rewrite  \tykl\
in the form of \ield\ (cf. \zetty)
 $$ Z (\b,F)  
=    {  V_{9} \b \ov  2 (2\pi)^{5} }
\int_{{\ep^2}}^{\infty } {d\s \ov \s^2 }\ \prod^{4}_{I=1} { \rf_I }\ 
    \bigg[
 \t_3 ( 0| {i\b^2 \ov  \pi \s})  \bigg(
 \prod^{4}_{J=1}{ \sinh  \hat \rf_J\s   \ov  \sinh  \rf_J\s } - {1\ov 2} \bigg) 
$$
\eqn\tyk{ \ \ \ \ \ \ \ \ - \ 
 \t_2 ( 0| {i\b^2 \ov  \pi \s})  \bigg(\prod^{4}_{J=1} { \cosh  \hat \rf_J\s   \ov  \sinh  \rf_J\s } - {1\ov 2} \bigg)  \bigg]
\  ,   }
or
$$
 Z (\b,F)  
=    {  V_{9} \b \ov  2 (2\pi)^{5} }
\int_{{\ep^2}}^{\infty } {d\s \ov \s^2 }\ 
    \bigg[
 \t_3 ( 0| {i\b^2 \ov  \pi \s}) \prod^{4}_{I=1} { \rf_I }  { \sinh  \hat \rf_I\s   \ov  \sinh  \rf_I\s } 
 - \ 
 \t_2 ( 0| {i\b^2 \ov  \pi \s}) \prod^{4}_{I=1} { \rf_I }  { \cosh  \hat \rf_I\s   \ov  \sinh  \rf_I\s } $$
\eqn\rerr{
   - \ { 1 \ov 2}  \t_4 ( 0| {i\b^2 \ov  4 \pi \s})  
  \prod^{4}_{I=1} { \rf_I }         \bigg]
\  .    }
In the zero-temperature limit $\b\to \infty$ 
the second term vanishes and we get back to \zetty,\tert.
In the zero-field limit $\rf_I \to 0$  it is the second $\t_2$-term 
that gives a  non-vanishing contribution  which is 
equal to the free energy of the  massless
$D=10$   SYM  modes  (cf. \zzee) 
 \eqn\tykm{  Z (\b,0)   
=   -  {  V_{9} \b \ov  2 (2\pi)^{5} }
\int_{0}^{\infty } {d\s \ov \s^6 }\ 
 \t_2 ( 0| {i\b^2 \ov  \pi \s})  = - 32  \k_9 V_9 \b^{-9}
\  .   }
The presence of the second $\t_2$-term
in \tyk\  with $\cosh  \hat \rf_I\s$ 
factors 
instead of $\sinh  \hat \rf_I\s$  in the first $\t_3$-term
can be related to  antiperiodicity  of the 
fermionic fields in euclidean time direction.
That  finite temperature 
 explicitly breaks supersymmetry is reflected   in the fact that 
in contrast to \zetty, the finite-temperature expression \tykl\ 
is non-trivial on  supersymmetric (e.g., self-dual) 
configurations with $\hat f_I=0$. 

In contrast to \zetty\  and the integral of the first $\t_3$-term  in \tykl, 
the integral  of the second $\t_2$-term  in \tykl\
is convergent for $t\to 0$
 (the 
finite  temperature provides an effective  UV cutoff in this term since 
$ \t_2  (0| {i\b^2 \ov \pi \s})_{\s\to 0}\  \to  \ 2 e^{-{\b^2 \ov 4\s}}$). 
Rescaling $\s$,  we can represent \tyk\ as
\eqn\repe{
Z(\b, F) =  \b^{-p}  H(\b^2 F) \ ,  }
$$
H(y) \equiv 
 a_0   \int^\infty_{\ep^2\ov \b^2} {d\s'\ov {\s'}^{p+3 \ov 2}} \  
\bigg[\t_3  (0| {i\ov \pi \s'}) \  G_3( \s' y)
- \t_2  (0| {i\ov \pi \s'}) \  G_2( \s' y)
- \t_4  (0| {i\ov 4\pi \s'}) \  G_4( \s' y) \bigg] \ , 
$$
where  we have added the factor  $\s^{(9-p)/2}$ to the measure in \tyk\
to describe the case of SYM theory in $D=p+1$ dimensions  and defined 
(cf. \nuuu) 
$$
G_3 (\s F) \equiv  \prod^{4}_{I=1} { \rf_I\s  }  { \sinh  \hat \rf_I\s   \ov  \sinh  \rf_I\s }
 = \ { 1 \ov 16} \s^4  \bigg[- 8 \prod_{I=1}^4 \rf_I  + 2
    \sum_{I=1}^4 \rf_I^4 -  (\sum_{I=1}^4 \rf_I^2)^2 \bigg]
  + O(\s^6\rf^6) \ , 
$$
$$
 G_2 ( \s  F) \equiv 
\prod_{I=1}^4  \rf_I\s\   { \cosh  \hat \rf_I\s  \ov  \sinh  \rf_I\s   }
=\ 1 + { 1\ov 3} \s^2 \sum_{I=1}^4 \rf_I^2 +  
{ 1\ov 720} \s^4 \bigg[ 39  \sum_{I=1}^4 \rf_I^4 - 10 (\sum_{I=1}^4 \rf_I^2)^2 \bigg]
  + O(\s^6\rf^6) \ , 
$$
$$
G_4 (\s F) \equiv { 1 \ov 2}  \s^4   \prod^{4}_{I=1} { \rf_I  }\  . $$
The weak-field expansion of $Z$ thus  has  the following structure
\eqn\expan{
Z   \ \sim\    b_1 \b^{-p } +  b_2 \b^{-p+4} F^2  + (b_0 + b_3 \b^{-p+8}) 
 F^4 + ... \ ,   }
where $b_0 F^4$ stands for  the $F^4 - \four (F^2)^2$ terms 
in the zero-temperature SYM effective action ($b_0\sim \ep^{8-p}$).  

For $t\to \infty$ in \tyk\ one finds the same  IR  singularity  as in the 
zero-temperature case. It is  known  that  finite temperature does not eliminate
the magnetic instability of the  YM theory  \cabo. 
Since this  instability  has its origin in the  vector-field sector, 
it is not affected  by the presence of fermions (irrespective of  the  
choice of their boundary conditions).

\subsec{\bf String theory  partition function in magnetic background}
Below we shall consider an ensemble of 
open superstrings  at finite temperature  in an external magnetic field.
Our aim 
 will be to  find  the finite temperature
 analogs of the partition functions \zete,\zetee,\zte.
 Taking the $\a'\to 0$ limit, we will reproduce  
the  free energy \tyk\  of the $N=4, \ D=4$ SYM theory  in a constant abelian
magnetic field. 

 As in the zero background field case  (cf. \refs{\pool,\alo}) 
we shall obtain  the 
finite temperature  string partition function 
$$Z(\b, F) = \b {\cal F}(\b, F) = - \ln \hat Z (\b,F)\ , $$ 
where $\b$ is the inverse temperature,  ${\cal F}$ is  the free energy
and $\hat Z$ is the canonical partition function of string field theory, 
using  the l.c. GS  path integral formalism.
Compared to the zero-temperature case  of 
section 2    now   the euclidean  time  coordinate 
 has  period $\b$   and thus includes  winding modes 
in the angular coordinate $\psi$ of the annulus.
The fermionic coordinate 
$S$ in \ope\  may be either periodic or   {\it antiperiodic}  in  $\psi$
(which  effectively 
plays the role of the euclidean time coordinate).
In fact, both  sectors  should be included in the GS partition function
with  appropriate  temperature factors
(this corresponds  to taking into account  different statistics 
in the  cases of 
 space-time  bosonic or fermionic states 
 propagating  in the loop).

The periodic sector  contribution
is the obvious generalisation of \zettt\  
\eqn\tett{
Z^+ (\b, F)=  a_1   \b \int^\infty_0 {dt\ov t^2}\ 
  \t_3  (0| {i\b^2 \ov 2\pi^2 \a' t}) 
\prod_{I=1}^4 \rf_I 
 { \t_1 (it\hb_I | it) \ov 
 \t_1 (it \vp_I | it) }  \  .    }
 It vanishes  in the absence 
of the external field but  provides 
 correspondence  with the 
zero-temperature expression for the  string  partition function \zettt\
in the limit $\b\to \infty$.

In the antiperiodic sector 
there  is no fermionic zero-mode factor 
 present in \zet,\tett, and 
 the space-time supersymmetry is explicitly broken by the temperature. 
The contribution $Z^-$ of the antiperiodic sector 
  can be written as  (cf. \zet)
\eqn\zett{
Z^- (\b,F) = -a_1   \ \int^1_0 {dq \ov q}\  {\cal Z} (\b, q) \ 
\prod_{I=1}^4  Z^-_I (\F, \FF; q) \ , 
\  }
where  $Z^-_I$  has the same form as in 
 \zza\ (and is equal to 1  for $F^{(r)}=0$),  
while the temperature-dependent factor  ${\cal Z} (\b, q)$ is the 
 same as in the absence of the magnetic field   \refs{\alo,\mac} 
\eqn\free{
 {\cal Z} (\b, q)
= \pi^4 \b \  \t_2  (0| {i\b^2 \tt\ov 2\pi^2 \a'}) 
\bigg[ { \t_4 (0| i\tau)  \ov \t_1' (0| i\tau)}\bigg]^4
  \ , \ \ \ \ \ \ \  q = e^{-\pi \tt}
    \ ,   }
 where 
$\t_1' (0| i\tau) = 2 \pi \eta^3 (i\tau)$. 
 Equivalent forms 
of ${\cal Z} (\b, q)$ are   found   using  that 
\eqn\xxx{
\t_2 ({iv\ov \tt}|{i\ov \tt}) 
= \sqrt {\tt}\  e^{\pi v^2 \ov \tt}\  \t_4(v| i\tau) \ , 
\ \ \ \ 
{\t_2 ({iv\ov \tt}|{i\ov \tt})  \ov
 \t_1 ({iv\ov \tt}|{i\ov \tt}) }
=   - i { \t_4(v| i\tau) \ov \t_1 (v|i\tau)} \ . }
To compute $Z^-$ we note that the  fermionic Green function $\hat K$ in  $Z^-_I$ 
has the same form as \func\ but now 
 with the sum going over half-integers $r=1/2,3/2,...$,  
 or,  equivalently,  over integers $n=r+\ha$,  but with 
$A_n \   \to\   \hat A_n = { 1 + q^{2n-1} \ov 1 - q^{2n-1}}.$
As a result, $Z^-_I$  becomes (cf.  \vtt) 
\eqn\vtte{
Z^-_I
= \prod_{n=1}^\infty { (1 - \hat f^{(1)}_I \hat f^{(2)}_I)^2 + 
(\hat  f^{(1)}_I + \hat f^{(2)}_I)^2 \hat A_n^2 \ov
(1 -  \f \ff)^2 + ( \f+  \ff)^2 A_n^2  }  \ , } 
i.e. (cf. \kik)
\eqn\kike{  
Z^-_I =\bigg([1+  (\f)^2][1+  (\ff)^2] \bigg)^{1/2}  
\prod_{n=1}^\infty { ( 1 - q^{2n})^2 \ov (1 - q^{2n-1})^2} 
\prod_{n=1}^\infty { 1 - 2 q^{2n-1} \cos 2\pi\tb_I  +  q^{4n-2} \ov
1 - 2 q^{2n} \cos 2\pi\vp_I  +  q^{4n} }\ . 
}
Here  we have used the fact  that  taking a constant out of the 
product  over half-integers does not produce an overall factor, 
$\prod_{r={1\ov 2}}^\infty c = 1$.\foot{The infinite    products
 are  to be regularised with the 
generalised $\zeta$-function  $\zeta(z,a) =  \sum^\infty_{n=0} (n+a)^{-z}$
  for which  $\zeta(0,0)=-  {1\ov 2}$ but    $\zeta(0,{1\ov 2})=0$. The latter relation is
  also the reason 
why the fermionic contribution (of the only possible antiperiodic 
spinor  on the disc) 
 does not change the bosonic  Born-Infeld  expression for the 
tree-level string effective action  \refs{\tse,\mrt}.}
Since  
\eqn\deft{
 \t_4 (\vp |i\tau) \equiv  \sum_{n=-\infty}^{\infty}
 (-1)^n e^{-\pi \tt n^2 + 2\pi n \vp}=  
 \prod_{n=1}^\infty (1-q^{2n})
 (1 - 2 q^{2n-1} \cos 2\pi \vp    +  q^{4n-2})  \ , }
we finally get (for the definitions of parameters  see \dee,\hatt,\fef)
\eqn\theeet{
Z^-_I = { \rf_I \ov  \pi} \ 
    {  \t_1'(0|i\tau)  \ov   \t_4 (0| i\tau) } \ 
 { \t_4 (\tb_I | i\tau) \ov 
 \t_1 (\vp_I | i\tau)}  \ .   } 
Combining this with \free\ we find that  $Z^-$ 
 \zett\ takes the following
 simple form  (cf. \zete)
\eqn\ana{
Z^-= -a_1 \b \int^\infty_0 {d\tau}\ 
  \t_2  (0| {i\b^2 \tt \ov 2\pi^2 \a'}) 
\prod_{I=1}^4 \rf_I \ 
 { \t_4 (\tb_I | i\tau) \ov 
 \t_1 (\vp_I | i\tau) }  \ . }
As in the zero-temperature case, 
to have the agreement with the NSR result one should
make the replacement of $\tb_I $ \hatt\ by $ \hb$  \deee\
in the fermionic $\t_4$-contribution (cf. \zetee) 
\eqn\anas{
Z^-= -a_1  \b \int^\infty_0 {d\tau}\ 
  \t_2  (0| {i\b^2 \tt \ov 2\pi^2 \a'}) 
\prod_{I=1}^4 \rf_I \ 
 { \t_4 (\hb_I | i\tau) \ov 
 \t_1 (\vp_I | i\tau) }  \ . }
This can be put  into  the form of an integral 
over   the open-string proper time variable  $t=1/\tt$ 
as in \zettt\foot{The exponential 
factors  appearing in the Jacobi transformation
again cancel out since $\sum^4_{I=1} (\hb^2_I - \vp^2_I) =0$.}
\eqn\antsa{
Z^-= - a_1  \b \int^\infty_0 {dt\ov t^2}\ 
  \t_2  (0| {i\b^2 \ov 2\pi^2 \a' t}) 
\prod_{I=1}^4 \rf_I 
 { \t_2 (it\hb_I | it) \ov 
 \t_1 (it \vp_I | it) }  \ .   }
$Z^-$ vanishes in the zero-temperature limit 
and reduces to the free string  partition function \alo\ 
in the zero-field limit. 

The final expression for the  partition function which 
is the finite-temperature analogue of \zettt\
is thus 
$$Z= Z^+ +   Z^- $$
\eqn\ansa{
= \ a_1  \b \int^\infty_0 {dt\ov t^2}\ 
\bigg[ \t_3  (0| {i\b^2 \ov 2\pi^2 \a' t}) 
\prod_{I=1}^4 \rf_I 
 { \t_1 (it\hb_I | it) \ov 
 \t_1 (it \vp_I | it) }  \ 
-  \  \t_2  (0| {i\b^2 \ov 2\pi^2 \a' t}) 
\prod_{I=1}^4 \rf_I 
 { \t_2 (it\hb_I | it) \ov 
 \t_1 (it \vp_I | it) }  \bigg]     .    }
The equivalent form of \ansa\ is 
$$
Z(\b, F) =  a_2  \int^\infty_0 {dt\ov t^{3/2}}\    \bigg[
  \t_3  (0| {2\pi^2 \a' it \ov \b^2 }) 
\prod_{I=1}^4 \rf_I\  
 { \t_2 (it\hb_I | it) \ov 
 \t_1 (it \vp_I | it) }
$$
\eqn\nsa{ \ \ \ \ \ \ \ \ \ \ \ \ \ \  -  \ 
\t_4  (0| {2\pi^2 \a' it \ov \b^2 }) 
\prod_{I=1}^4 \rf_I\  
 { \t_2 (it\hb_I | it) \ov 
 \t_1 (it \vp_I | it) } \bigg] \ ,}
where (cf. \coff) 
\eqn\norm{
a_2 = (2\pi^2 \a')^{1/2}  a_1 \ , \ \ \  \ \ \   
a_1  =   { N^2  V_9\ov 2 (2\pi)^5 (2\pi \a')^5}  \ ,  
}
and $N$ is the Chan-Paton number (number of D9-branes). 
Using  a version of the  Riemann identity \jaco\
$Z$  can be   written  also  in  the NSR  form similar 
to \ztep,\zte\ (cf.  \atic). 

\subsec{\bf Some properties of  finite temperature partition function}
Starting with the string-theory partition function \ansa\  
and repeating the same steps 
as in section 2.4, i.e.
restoring the $2\pi \a'$ factors in $f_I$, defining $\s\equiv  2\pi \a' t$ and 
taking the $\a'\to 0$ limit,   one finds indeed 
the SYM partition function \rerr\ (apart from the last $\prod^4_{I=1} \rf_I$ term
the role of which in the zero-temperature 
 field-theory expression is to cancel a similar
product term coming out of the expansion of the first term in \rerr, see 
 section 2).

In the case of the  `neutral' background  ($\rf_I=\f+\ff=0$)   the partition 
function   $Z(\b,F)$  is equal to $Z^-$ and reduces 
 (as  in the bosonic string theory) simply 
to its 
  zero-field  value $Z(\b,0)=-a_1 \int{dq\ov q}  {\cal Z} (\b,q)$ 
multiplied by the overall 
factor $\det(\d_{ij}+ F_{ij}) = \prod_{I=1}^4 ( 1 + f^2_I)$
(cf.  \kike). 

The behaviour of the integral \nsa\ in the open-string  UV region 
$t\to 0$  is determined by the $\tt\to \infty $ region
  of \anas: 
since  $\t_4(\hat \vp_I|i\tau)\to 1, $
$\t_1( \vp_I|i\tau)\to 2 e^{- {\pi \tt\ov 4}} \sin \pi \vp_I , $\ 
$\t_2 (0| {i\b^2 \tt \ov 2\pi^2 \a'}) \to  2 e^{- {\b^2 \tt \ov 8\pi \a'}}$, 
we conclude that the integral is convergent at $t\to 0 $   provided  
 $\b > \b_c = 2\pi \sqrt{2 \a'}$. 
Thus the presence of the {\it magnetic} field does not  change 
the  value   of the  Hagedorn  temperature 
of the free  open superstring gas. 

The open-string  IR behaviour  $t\to \infty$  of the integral \nsa\ 
is similar to that 
in the zero-temperature case \zettt,\simi\ discussed in sect. 2.2
(the temperature-dependent  factors in \nsa\ 
 become  trivial in this limit, 
$\t_{3,4} (0|{2\pi^2 \a' i t \ov \b^2})|_{t\to \infty} \to 1$), 
\eqn\siml{
Z\ \to \  \int^{t\to \infty} {dt\ov t^{3/2}} \  \bigg[
\prod_{I=1}^4  \rf_I\  { \sinh \pi t  \hat  \vp_I  \ov  \sinh  \pi t  \vp_I 
   }  - 
\prod_{I=1}^4  \rf_I\  { \cosh \pi t  \hat  \vp_I  \ov  \sinh  \pi t  \vp_I 
   } \bigg] \      }
$$
= - { 1 \ov 4}  \int^{t\to \infty} {dt\ov t^{3/2}} \  
\prod_{I=1}^4 {   \rf_I\    \ov  \sinh  \pi t  \vp_I 
   } \sum^4_{J=1} \cosh 2 \pi t  \hat  \vp_I  \   
, $$
where we have used \uyy,\uyyu. 
The conditions of convergence of this integral are the same as  for  
\simi, i.e.  one finds  the same magnetic IR instability
as in the zero-temperature case.

The partition function \nsa\  gives free energy of the open string gas on D9-brane. 
The Dp-brane version of \anas\ is obtained by T-duality
(or by taking  limits when some of $f^{(r)}_I$ go to zero or infinity)
as in the zero-temperature case
discussed in section 2.3.
The case when $\f=\ff=f_I$ corresponds to 
 the  gas of open strings with both ends attached to the same 
Dp-brane,
 so that $Z$ may be interpreted as determining 
the  thermal self-energy of the 
Dp-brane (or  correction to the tension  of a Dp-brane in a thermal state).
 The temperature should   correspond to the  Hawking temperature 
of a 
 non-extremal brane in the   supergravity description, while the 
magnetic field  may be used to  represent    bound states 
of D-branes.

Keeping $\f$ and $\ff$ general  and adding the factor 
$e^{-{r^2 \ov 2\pi \a'}t}$ one finds the  thermal 
partition function of open strings stretched between two Dp-branes.
 The zero magnetic field  limit of  the Dp-brane  partition function is
 \refs{\greee,\vaz}
\eqn\fee{
Z(\b,0) = -{ 8\pi^4 N^2 V_p\ov  2 (8 \pi^2 \a')^{p/2} } 
\int^\infty_0{ dt\ov t^{1+{1\ov 2} p}} \ e^{-{r^2 \ov 2\pi \a'}t}  
 \ \t_4 (0|{2\pi^2i \a' t \ov \b^2}) \ 
  \bigg[{\t_2 (0|it) \ov \t_1'(0|it)}\bigg]^4  \ , 
}
where $
8\pi^4 \bigg[{\t_2 (0|it) \ov \t_1'(0|it)}\bigg]^4
= 8
\prod^\infty_{n=1} \bigg({ 1 + e^{-2\pi t n } \ov  
  1 - e^{-2\pi t n } } \bigg)^8 .$
The $p=9$  case of \fee\ 
corresponds to   the $F^{(r)}=0$ limit of \nsa. 
For example, the  finite-temperature analogue of the D8-brane expression \ptt\
is given by the obvious modification of \nsa\ which reduces to \fee\ for $F=0$.

Having in mind  possible applications to the study of potentials between 
extremal and non-extremal branes one  would be  interested  in 
expanding  $Z(\b,F)$ for small  field  and  temperature.
In contrast  with  the zero-temperature case where the expansion of $Z$ 
in powers of $F$ starts  with the universal $F^4$ term, here 
one  finds  also   lower powers of $F$ with temperature-dependent 
coefficients, 
 and there is no a priori reason for the  `universality'
of such terms. 
 The leading terms  in the  expansion of the partition function \nsa\
in powers of the field  will have  the form  (cf. \expan) 
$\ 
Z  \ 
\sim\    k_1 \b^{-9} +  k_2 \b^{-5} F^2  + (k_0 + k_3 \b^{-1})  F^4 + ... , $
where $k_i$ with $i > 0$ will  be non-trivial 
 functions of $r$ and $\a'$.

\subsec{\bf Neutral superstrings in  electric background}

In general, one  does not expect an equilibrium 
distribution for charged strings in 
 an electric field, 
so it is natural to consider the neutral string case,
$f^{(1)}_0 =-f^{(2)}_0 =  i E, \ E= 2\pi\a'\E$.
The corresponding partition function $Z(\b,E)$ can be computed  
using either real-time or imaginary-time formalism. 
For the bosonic string $Z(\b,E)$  was found in \ffi. 
In the real-time approach  one should  take into account  that 
according to \nester\
 the oscillation part of the mass spectrum of a bosonic 
 neutral string in electric  and magnetic fields 
  gets  rescaled by  $1-E^2$ factor  
 \eqn\fcc{M^2 = -\sum^{D-2\ov 2}_{I=1}  { E^2 + f^2_I \ov 1 + f^2_I}\ P_I^2 
 + {1\ov \a'} (1-E^2) \sum^\infty_{n=1} \sum^{D-2}_{i=1} n \bar a^i_n  a^i_n \ ,  }
 where the oscillators are  canonically 
 normalised,  $[a^i_n, \bar a^j_m]= i\d^{ij}\d_{nm}$.\foot{The  special role of the electric field compared to the 
 magnetic one is
 related to the fact that it couples to the time 
   coordinate and thus contributes to the momentum.
 The definition of momentum becomes important at non-zero 
 temperature since the partition function is defined by a phase-space integral.}
 Using the open string theory 
 proper-time representation one then finds  that 
 the  factor $1-E^2$ appears  multiplying the integration variable
 $t$ in the arguments of $\t$-functions,  or multiplying
 the temperature term  $\b^2/t$   after a redefinition of $t$. 
 In the imaginary-time approach one arrives at the same 
 expression after taking into account that the winding modes 
 of  the imaginary time coordinate couple to the electric field 
 term in the string  action.  As a result, the critical temperature 
 of the bosonic string gets rescaled by the  factor   $\sqrt{1-E^2}$
 \ffi.
 
 Similar conclusion is  reached in the superstring case
 (for simplicity we shall set the magnetic components 
 to  zero  since  the generalisation to  the `mixed-field' case is obvious).
In the case of the neutral string in the background field with 
just one electric component 
we get 
\eqn\aisa{
Z=   - a_1 \pi^{4} [1 - (2\pi\a'\E)^2]\ \b \int^\infty_0 {dt\ov t^6}\ 
  \t_2  \big(0| {i  [1 - (2\pi\a'\E)^2]\b^2 \ov 2\pi^2 \a' t}\big) \  
 \bigg[{ \t_2 (0 | it) \ov \t'_1 (0 | it) }\bigg]^4  \ .   }
 As follows from \aisa,
 the  critical inverse temperature ($\b > \b_c$) 
 becomes dependent on the electric field:
 \eqn\criti{
 \b_c=   {2\pi \sqrt{2\a'}
 \ov  \sqrt{1-(2\pi\a'\E)^2 }}\ . }
 The  field-dependent rescaling  factor may be interpreted as a 
 modification ($T_0\to T_{\rm eff}$) of the string tension 
 which enters the expression for the critical temperature,\foot{This 
 rescaling of the open-string tension is similar to the one discussed
 in D-string context in \guk.} 
\eqn\tenn{
T_{\rm eff} = T_0 [1- (T^{-1}_0\E)^2]  
\ , \ \ \ \ \ \ \ \ T_0 = {1 \ov 2\pi\a'} \ . 
}
Note that the SYM  theory  limit of \aisa\  is trivial 
as  all dependence on the electric field disappears  
 for   $\a'\to 0$.\foot{For  discussions  of related 
 (electric field, finite temperature) problems in field theory
  context see, e.g., \elm, where 
the  
one-loop effective action of  Dirac fermions 
 in an approximately constant electro-magnetic field 
 at finite temperature  was  computed,  and \chap, where 
 a weak-field expansion of the one-loop 
 finite temperature YM effective action in
  generic background  was considered.
 }

\bigskip\bigskip

\centerline {\ \bf Acknowledgments}
\bigskip
I am  grateful to I. Chepelev, T. Evans, M. Gutperle, H. Liu, R. Siebelink 
and V. Zeitlin
for useful discussions 
 and collaboration on  some related issues. 
 I  acknowledge  also the support
 of PPARC and  the European
Commission TMR programme grant ERBFMRX-CT96-0045.
\vfill\eject
\listrefs
\end